\documentclass[fleqn,usenatbib]{mnras}

\usepackage[T1]{fontenc}
\usepackage{ae,aecompl}

\usepackage{graphicx}	% Including figure files
\usepackage{amsmath}	% Advanced maths commands
\usepackage{amssymb}	% Extra maths symbols
\usepackage{gensymb}
\usepackage[table,xcdraw]{xcolor}
\usepackage{graphicx}
\usepackage{indentfirst} % automatic indentation of paragraphs 
\usepackage{lipsum}  % insert dummy text for testing
\usepackage{enumerate}
\usepackage{enumitem}
\newcommand{\pl}[1]{\textcolor{red}{\bf Paulo: }}
\newcommand{\db}[1]{\textcolor{blue}{\bf Douglas: }}

\usepackage{newtxtext,newtxmath}

\title[The role of clusters in galaxy quenching]{Examining transitional galaxies to understand the role of clusters and their dynamical status in galaxy quenching}

\author[Brambila et al.]{
Douglas Brambila$^{1}$,
Paulo A. A. Lopes,$^{1}$, André L. B. Ribeiro$^{2}$, Arianna Cortesi$^{1}$
\\
$^{1}$Observatório do Valongo, Universidade Federal do Rio de Janeiro, Ladeira do Pedro Antônio 43, Saúde, \\ Rio de Janeiro, RJ 20080-090, Brazil -- douglas13@ov.ufrj.br \\
$^{2}$Laboratório de Astrofísica Teórica e Observacional -- Departamento de Ciências Exatas e Tecnológicas --Universidade Estadual de Santa Cruz,\\ 45650-000, Ilhéus, BA, Brazil\\
}

\date{Accepted XXX. Received YYY; in original form ZZZ}

\pubyear{2022}

\begin{document}
\label{firstpage}
\pagerange{\pageref{firstpage}--\pageref{lastpage}}
\maketitle

\begin{abstract}

In this work, we consider four different galaxy populations and two distinct global environments in the local Universe (z $\leq 0.11$) to investigate the evolution of transitional galaxies (such as star-forming spheroids and passive discs) across different environments. Our sample is composed of 3,899 galaxies within the R$_{200}$ radius of 231 clusters and 11,460 field galaxies. We also investigate the impact of the cluster's dynamic state, as well as the galaxy's location in the projected phase space diagram (PPS). We found that although the cluster environment as a whole influences galaxy evolution, the cluster dynamical state does not. Furthermore, star-forming galaxies represent recent cluster arrivals in comparison to passive galaxies (especially in the case of early-types). Among the ETGs, we find that the D$_n(4000)$ and H$_\delta$ parameters indicate a smooth transition between the subpopulations. In particular, for the SF-ETGs, we detect a significant difference between field and cluster galaxies, as a function of stellar mass, for objects with Log $M_*$/M$_{\odot} > 10.5$. Analyzing the color gradient, the results point toward a picture where field galaxies are more likely to follow the monolithic scenario, while the cluster galaxies the hierarchical scenario. In particular, if we split the ETGs into lenticulars and ellipticals, we find that the steeper color gradients are more common for the lenticulars. Finally, our results indicate the need for galaxy pre-processing in smaller groups, before entering clusters.

\end{abstract}


%%%%%%%%%%%%%%%%%%%%%%%%%%%%%%%%%%%%%%%%%%%%%%%%%%

%%%%%%%%%%%%%%%%% BODY OF PAPER %%%%%%%%%%%%%%%%%%

%------------------------------------------------------------------------------------------------------------------%
%------------------------------------------------------------------------------------------------------------------%
%------------------------------------------------------------------------------------------------------------------%
\section{Introduction}
\label{sec:intro}
% This is a simple template for authors to write new MNRAS papers.
% See \texttt{mnras\_sample.tex} for a more complex example, and \texttt{mnras\_guide.tex}
% for a full user guide.

% All papers should start with an Introduction section, which sets the work
% in context cites relevant earlier studies in the field by \citet{Others2013},
% and describes the problem the authors aim to solve \citep[e.g.][]{Author2012}.

The existence of a bi-modality in different galaxy properties is well established in the past years  \citep{strateva+01, kau03b, baldry+04, balogh+04, baldry+06}. We can separate galaxies into two main categories: blue cloud galaxies (BC) -- mostly blue star-forming spiral galaxies; and red sequence objects (RS) --- mainly red early-types with no significant star-formation activity. Despite that, we have found galaxies that defy the simplistic view that elliptical galaxies are red and dead while spiral galaxies are blue and lively (e.g., \citealt{wolf+09, masters+10, cortese12, crossett+14, lopes+16, kuc17, mah18}). We have also established the existence of an intermediary region between the BC and the RS, called Green Valley (GV), that works as a transitioning region between the BC and RS \citep{martin+07, salim+09, mendez+11, schawinski+14}. These two points together paint a clear scenario where galaxies evolve from the BC to the RS as their life progresses. However, the mechanisms responsible for this transformation are still open for debate.

\citet[][]{dressler1980} showed that there is a relation between the galaxy morphology and the environment where the galaxy inhabits (the Morphology-Density Relation). Early-type galaxies (ETGs; the dominant morphology in the RS) show a preference for inhabiting higher-density environments, and therefore are primarily found in galaxy groups or clusters. On the contrary, late-type galaxies (LTGs; the dominant morphology in the BC) show a preference for inhabiting lower-density environments, being mainly found in the field or the outskirts of groups and clusters. Alongside that relation, we also have evidence that the fraction of blue galaxies inside clusters increases while the fraction of red galaxies decreases when we go for higher redshift regimes \citep{butcher&oemler84}.

Different mechanisms impact a galaxy as it enters a cluster. Those can be responsible for quenching the galaxy's star formation or, at least, accelerating this process. One of those effects is ram-pressure striping (RPS), where the gas inside the galaxy is removed due to the pressure caused by the hot intracluster medium \citep[][]{gunn&gott-72}. Other ones are strangulation and starvation --- when the hot intracluster medium removes gas from the galaxy halo and also prevents a new supply of gas from falling into the galaxy \citep[][]{larson+80}. We also have events like mergers (when two galaxies go into a fusion process and become one single entity --- \citealt{toomre&toomre-72}), harassment (the accumulative effect of several fly-by encounters of a galaxy with others --- \citealt{moore+96, moore+98}), and tidal force from the gravitational potential well of the cluster (capable of driving gas to the center of the galaxy and triggering a center starburst and bar instabilities --- \citealt{byrd&valtonen-01, lokas+16}). All of those processes are, in some way or another, capable of altering the gas content of a galaxy inside or entering the cluster.

On the other hand, one cannot ignore that galaxies also evolve when found in isolation in the Universe. This isolated evolution points toward a myriad of processes happening inside those galaxies. This event falls under the umbrella of mass quenching. In lower mass systems (M$_* < 10^9 M_\odot$), stellar and AGN feedback are capable of heating and, in some cases, even removing galactic gas (e.g., \citealt{ponman+99}), while for higher mass systems, the importance of stellar feedback decreases, but the AGN one is still very important \citep[][]{larson74,vecchia+08, coton+06, fabian+12}.

Several studies show that mass quenching and environmental quenching have distinct effects on different galaxy types. According to \citet{peng10}, satellite galaxies are more likely to be quenched by environmental action, whereas central galaxies are more likely to be quenched by mass-related effects.

It is not an easy task to separate the influence of internal and external mechanisms in galaxy evolution. Besides showing that mass and environmental quenching have a distinct impact on different types of galaxies, \citet{peng10} also verified that environmental quenching acts independently of the mass and vice-versa. 

A clear evidence of the cluster influence in galaxy evolution is presented by \citet{Jaffe+15}. The authors analyzed galaxies of the relaxed cluster Abell $963$ (z $= 0.203$) observed with the Blind Ultra Deep HI Environmental Survey (BUDHIES). They showed that galaxies lose their gas content, down to the BUDHIES detection limit, in their first passage into the cluster center. The authors also encounter a significant fraction of galaxies that arrive at the cluster already without gas content. This last result raises the question of the role of the cluster in the gas removal of those galaxies.

In the present work, we try to shed some light on the role of clusters in galaxy quenching, investigating the properties of different galaxy populations and at different locations within clusters. We focus on the comparison of transitional galaxies - such as the star-forming spheroids and red discs - to the more regular populations. That is done in different locations in the projected phase space diagram. We also investigate possible dependencies to the cluster's dynamical state and compare the cluster results to what is found in the field. 

The paper is organized as follows: In $\S 2$, we introduce our data, describing the galaxy morphological and star-formation activity classification, as well as the cluster dynamical state, while in $\S 3$, we show our results. In$\S 4$, we have a discussion of the main results and our conclusions. The cosmology assumed in this work is $\Omega_{\rm m}=  0.3$, $\Omega_{\lambda}=  0.7$, and H$_0 = 100$ $\rm h$ $\rm km$ $s^{-1}$ Mpc$^{-1}$, with $\rm h$ set to $0.7$.
%------------------------------------------------------------------------------------------------------------------%
%------------------------------------------------------------------------------------------------------------------%
%------------------------------------------------------------------------------------------------------------------%

\section{Data}
\label{sec:data}

In this work, we consider galaxies from two very different environments, as we have a sample of cluster galaxies and one of field objects. The cluster sample is a combination of objects selected in different wavelengths. We have clusters from the supplement version of the Northern Sky Optical Cluster Survey (NoSOCS, \citealt[][]{Lopes+04, lop09a}), the Cluster Infall Regions in the SDSS (CIRS, \citealt{rin06}), the HIghest X-ray FLUx Galaxy Cluster Sample (HIFLUGCS, \citealt{rei02, Andrade-Santos+17}), the Planck Early Sunyaev-Zel'Dovich (ESZ, \citealt{pla11}) and the SPIDERS catalog \citep{kir21}.

%------------------------------------------------------------------------------------------------------------------%
\begin{figure*}
    \includegraphics[width=\linewidth]{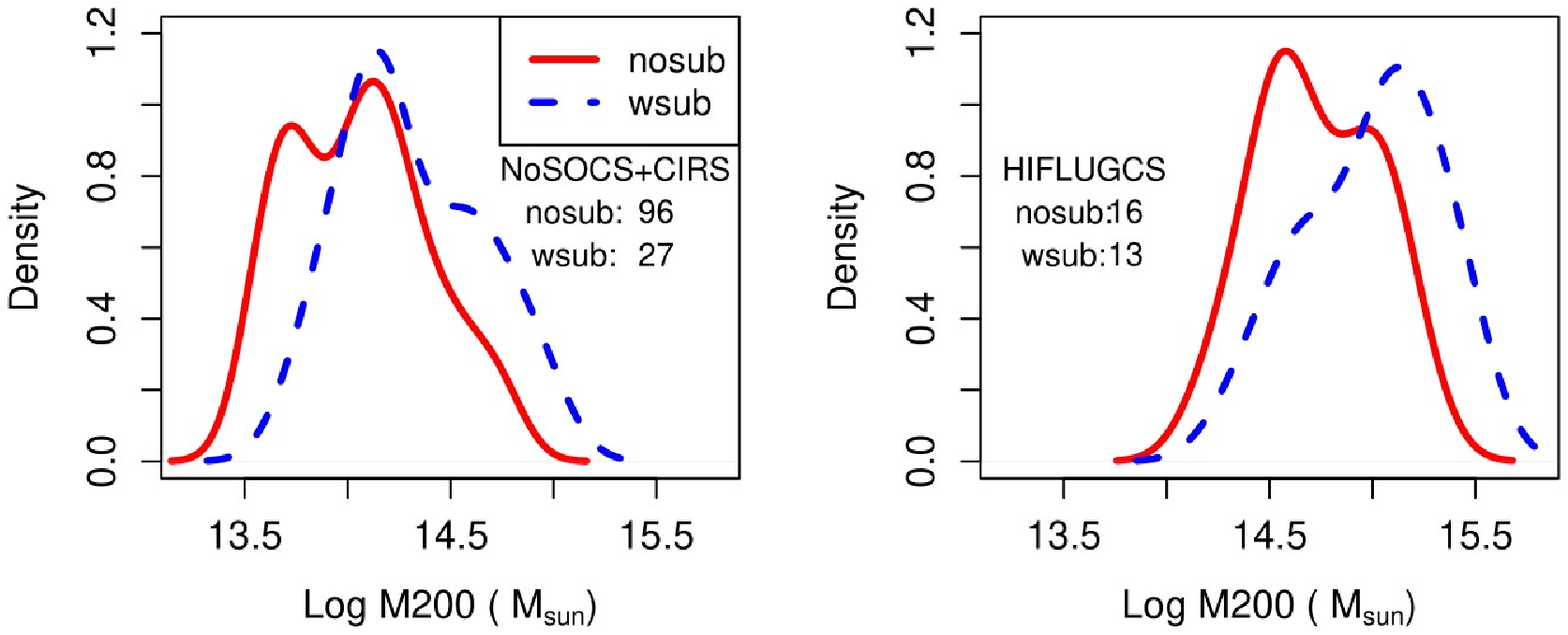}
    \includegraphics[width=\linewidth]{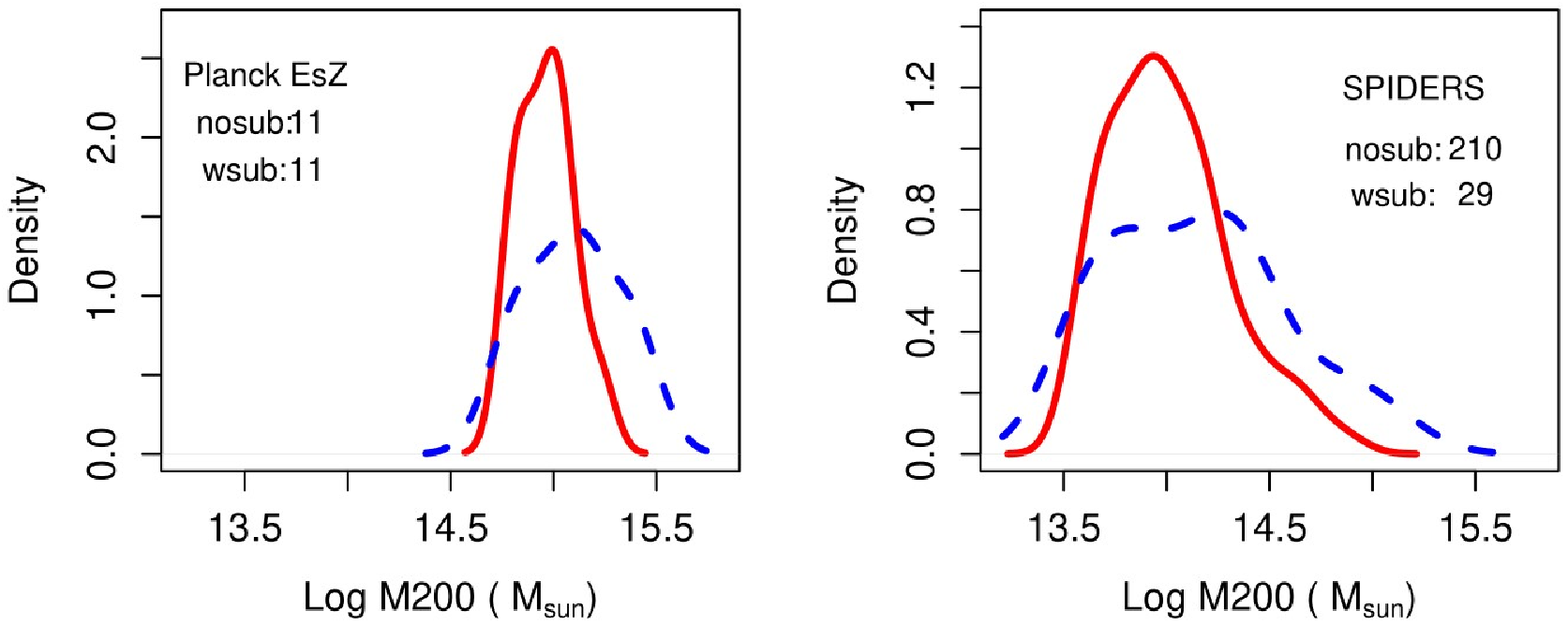}
    \caption{The distribution of Log M$_{200}$ for the cluster samples considered for the present work. In the top left we show the NoSOCS+CIRS clusters, in the top right the HIFLUGCS objects are displayed, while in the bottom we have the Planck ESZ (left) and SPIDERS (right) samples. The distributions are for the original data sets, before we check for duplicates and remove them, but we do consider the redshift and mass cuts described above ($z \le 0.10$ and LOG M$_{200}/M_\odot \ge 13.5$). On each panel, we display objects with no substructure in red (solid lines) and those with substructure in blue (dashed lines). The number of systems with or without substructure is also listed in the panels.}
    \label{fig:mass_distri_clusters}
\end{figure*}
%------------------------------------------------------------------------------------------------------------------%

The NoSOCS clusters comprise the only optically selected catalog of those listed above. HIFLUGCS is an X-ray cluster catalog. The CIRS and SPIDERS samples are composed of X-ray selected clusters, with SDSS spectroscopic data. The latter is actually based on an extensive follow-up effort of SDSS-IV. The Planck ESZ catalog consists of clusters selected through the Sunyaev-Zel'Dovich effect. All the clusters we use in the current work are within the SDSS DR7 spectroscopic footprint, so that we can uniformly derive cluster properties for these systems. We have previously worked with the NoSOCS+CIRS clusters in \citet{lop09a, lop09b, Lopes+14, lopes+16, lop17} and \citet{rib13}. In \citet{Lopes+18}, we compared substructure estimates and BCG properties of those clusters to the ones from the HIFLUGCS and Planck ESZ lists (as provided by \citealt{Andrade-Santos+17}). In the current paper, we combined all these catalogs with the SPIDERS sample. Our goal is to minimize possible biases due to different selection techniques and wavelengths. One bias example is related to the substructure fraction, discussed in \citet{Andrade-Santos+17} and \citet{Lopes+18}, that can be found when comparing X-ray and SZ selected samples (also seen when comparing to optically selected systems). However, in order to have a more uniform data set we only consider systems at $z \le 0.10$, containing at least 10 members within R$_{200}$ and with LOG M$_{200}/M_\odot \ge 13.5$ (membership, R$_{200}$ and M$_{200}$ estimates are explained below). Doing so, we are able to work with a complete spectroscopic sample from the SDSS DR7 main sample and derive reliable estimates of membership and cluster physical parameters.

Figure \ref{fig:mass_distri_clusters} shows the distribution of Log M$_{200}$ for each cluster sample mentioned in the previous paragraph. The number of objects with (dashed line) or without (solid line) substructure is also indicated. The distributions are for the original data sets, before we check for duplicates and remove them (see below), but we do consider the redshift and mass cuts described above ($z \le 0.10$ and LOG M$_{200}/M_\odot \ge 13.5$). This figure shows that the combination of the samples is an important step towards a more complete distribution according to the cluster mass. We are also able to mitigate the impact of cluster catalogs heavily affected by substructure (such as the Planck ESZ list). Note that systems with substructure generally have higher masses in comparison to the more regular clusters, a result consistent to the findings of \citet{rib11}.

All the clusters in our sample have the same redshift limits and had a minimum number of member galaxies and cluster mass imposed. However, the main factor that could still affect our results is the cluster mass cut, as we also consider groups, or low mass clusters. We verified that applying a higher mass cut (at LOG M$_{200}/M_\odot = 14$, instead of $13.5$) do not affect our main results. For instance, the fractions of the galaxy populations displayed in Table \ref{tab:tabela_1} (see below) do not show a large variation. Note also, that the fraction of all galaxies within the groups (LOG M$_{200}/M_\odot < 14$) is small ($23\%$). Hence, we decided to keep the analysis with all systems in our sample, as described above (with LOG M$_{200}/M_\odot \ge 13.5$).

For each cluster from the above catalogs, we used SDSS-DR7 photometric and spectroscopic data to select members (and exclude interlopers), estimate the velocity dispersion ($\sigma_{cl}$), physical radius (R$_{500}$ and R$_{200}$) and mass (M$_{500}$ and M$_{200}$). 

As in \citet{lop09a} we select members and exclude interlopers after applying the shifting gapper technique \citep{Fadda+96} to all galaxies with available redshifts around each cluster. We only use the members within $2.5$ h$^{-1}$ Mpc to derive an initial estimate of the velocity dispersion ($\sigma_{cl}$). Differently from what we did in previous works when we estimated physical radius and cluster masses from a virial analysis (following the approach of \citealt{gir98}), we now follow the procedure described in \citet{fer20}. First, we apply the corrections proposed by \citet{fer20} to the velocity dispersion estimate (initially derived within $2.5$ h$^{-1}$ Mpc). Next, we obtain an estimate of M$_{200}$ adopting the equation 1 (listed below) of \citet{fer20} (also see \citealt{mun13}). The corrections considered by \citet{fer20} to the mass estimate are also employed.

\begin{equation}
    \frac{\sigma_{{\rm 1D}}}{{\rm km ~s^{-1}}} = A \left[ \frac{h(z) M_{200}}{10^{15} ~{\rm M\odot}} \right]^{\alpha},
    \label{eq:eq1_fer20}
\end{equation}
where $A$ is $1777.0$ ${\rm km ~s^{-1}}$ and $\alpha = 0.364$.

R$_{200}$ is assumed to be the radius at which the averaged mass density reached $200$ times the critical mass density of the Universe at redshift $z$. Hence, considering the mass $M_{200} = (4\pi/3)200 \rho_c(z) R^3_{200}$ is the total mass within R$_{200}$, we can derive an estimate of R$_{200}$ from the above mass estimate. We then obtain a final mass estimate, but now considering only members within R$_{200}$ (instead of $2.5$ h$^{-1}$ Mpc). Having this updated member list we estimate again $\sigma_{cl}$ and the mass, using equation 1 of \citet{fer20, mun13}. The corrections from \citet{fer20} are again applied both to $\sigma_{cl}$ and M$_{200}$. R$_{500}$ and M$_{500}$ estimates are derived after assuming a NFW profile and interpolating to the appropriate radius. For more details on the estimates above, we refer the reader to \citet{lop09a, Lopes+14, Lopes+18} and \citet{fer20}.

Next, we eliminate common clusters from the different catalogs, arriving at a cluster sample with 231 objects. They are between $0.018 \leq z \leq 0.100$, and have $13.5 \le $ LOG M$_{200}/M_\odot \le 15.13$. In order to keep a cluster, we also impose a minimum number of 10 members within R$_{200}$. The upper redshift limit ($z = 0.100$) corresponds to the completeness limit of the SDSS main spectroscopic sample, limited at $r_r^{petro} = 17.77$. That is translated to an absolute magnitude limit of $M_r \sim M^* + 1 = -20.58$. We call galaxies more luminous than this absolute magnitude limit as bright, which are the ones studied in the current work. In total, we have $3899$ bright galaxies ($M_r \le M^* + 1$), inside R$_{200}$ of those 231 galaxy clusters, ranging between $0.015 \leq z \leq 0.107$ and $10^{9.0} \leq M_*/M_\odot \leq 10^{12}$. For all those galaxies, we have morphological information from \citet[][hereafter DS18]{Dominguez-Sanchez+18} or \citet[hereafter HC11]{Huertas-Company+11}; see $\S$\ref{subsec:morph_class}. Although most of our work based on the cluster sample considers only galaxies within R$_{200}$, we still have member galaxies up to 5$\times R_{200}$ (approximately the turn-around radius). In total, we have $9931$ members within 5$\times R_{200}$ (which are used in Fig. \ref{fig:sf_frac}). It's important to notice that our sample includes low-mass galaxy systems (M$_{200} < 1\times10^{14}$), which should be referred to as groups. However, for ease of reading, we decided to call, in this paper, all the objects in our sample as clusters. Groups correspond to $\sim 40\%$ of our sample, containing $\sim 23\%$ of all the galaxies in our {\it cluster} sample.

The field sample was selected by comparing the  photometric and spectroscopic SDSS-DR7 data with a sample of about $15,000$ groups and clusters provided by \cite{gal09}, combined with our cluster sample described above. To select field galaxies, we excluded any galaxy with a distance smaller than 4 Mpc and with $\Delta z < 0.06$ of any object from the combined cluster catalog. To ensure a lower level of contamination from clusters and groups that eventually are not present in the cluster comparison sample, we also remove galaxies with a local galaxy density LOG$ (\Sigma_5) < -0.5$. The parameter $\Sigma_5$ (described in \citealt{lab10, Lopes+14, lop17}) is given by $5/\pi d^2_n$, where $d_n$ was selected to be the projected distance of the fifth-nearest galaxy with maximum velocity offset of $1,000~km s^{-1}$. The total number of field galaxies selected according to the criteria above is $55,842$ sources. Our sample was reduced to $\sim 30,000$ galaxies after we impose redshift ($0.01 \leq z \leq 0.11$), stellar mass ($10^{9.0} \leq M_*/M_\odot \leq 10^{12}$) and luminosity cuts ($M_r < M^* + 1$). Note that we also require the galaxies to have morphological information provided by DS18 (or HC11). Finally, we construct a sample of field galaxies that follows the mass distribution of the cluster sample. To do so, we select a random sample (of 3899 galaxies) from the $30,000$ field galaxies, forcing it to have the same pattern of stellar mass distribution that the cluster sample has. We then apply a Kolmogorov–Smirnov (KS) and Anderson-Darling (AD) test to check if the samples are statistically similar. In this case, we keep the selection and repeat the process, removing the already selected galaxies from the $30,000$ pile. We repeat the process until the KS and AD return that the sample is statistically distinct. In the end, we retain the largest sample of field galaxies that has a similar stellar mass distribution to the cluster sample, resulting in $11,460$ field objects. The $p-value$ is 0.13 and the AD test has the confidence of $99.5\%$.
%------------------------------------------------------------------------------------------------------------------%
%------------------------------------------------------------------------------------------------------------------%
\subsection{Galaxy morphology}
\label{subsec:morph_class}
Galaxy morphology was mainly extracted from DS18. They adopt a deep learning technique to classify galaxies from the sample, described at \cite{meert+15, meert+16}. This data set was constructed from the SDSS-DR7 spectroscopic main sample, taking into account redshift ($0.005 < z < 1.0$) and magnitude ($14 < m_r^{petro} < 17.77$) cuts. DS18 provides morphological information for about 670,000 galaxies. The Convolutional Neural Network was trained to obtain the T-type values and the probability of the galaxies containing certain features like a disc, inclination (face/edge-on), bar signature, bulge prominence, roundness, and mergers. They also provide the probability of a galaxy having a lenticular morphology.

To create a robust classification, we combine the T-type information and the probabilities provided by DS18 to have a  morphological classification for elliptical and lenticular galaxies. We proceed as below:

%------------------------------------------------------------------------------------------------------------------%
\begin{itemize}[align=left, itemindent=1em, labelsep=0em, labelwidth=1em, leftmargin=0pt, noitemsep]
\item \textbf{Elliptical (E): }
    \begin{itemize}[align=left, itemindent=2em, labelsep=0em, labelwidth=1em, leftmargin=0pt, noitemsep]
        \item T-type $< - 1.5$
        \item P$_{bulge} > 0.6$
        \item P$_{disk} < 0.8$
    \end{itemize}
\end{itemize}
%------------------------------------------------------------------------------------------------------------------%

%------------------------------------------------------------------------------------------------------------------%
\begin{itemize}[align=left, itemindent=1em, labelsep=0em, labelwidth=1em, leftmargin=0pt, noitemsep]
    \item \textbf{Lenticular (S0):}
    \begin{itemize}[align=left, itemindent=2em, labelsep=0em, labelwidth=1em, leftmargin=0pt, noitemsep]
        \item $-1.5 <$ T-type $< 0.0$
        \item P$_{S0} > 0.6$ 
        \item P$_{disk} < 0.8$
    \end{itemize}
\end{itemize}
%------------------------------------------------------------------------------------------------------------------%

%------------------------------------------------------------------------------------------------------------------%
 \begin{table}
 \resizebox{\columnwidth}{!}{%
 	\centering
 	\begin{tabular}{cccccc} % four columns, alignment for each
 		                  \multicolumn{6}{c}{Cluster sample} \\
 		                  & \textbf{Passive}    & \textbf{GV} & \textbf{SF}       & \textbf{GV+SF} & \textbf{Total}\\ \hline
 		\textbf{ETG}    & 2432                &  77         & 56                & 133            & 2565\\
 		                  & (94.81\%)           &  (3.00\%)   & (2.18\%)          & (5.18\%)       & (65.79\%)\\ 
            Ellipticals (E) & 2003                & 39          & 24                & 63             & 2066\\ 
            Lenticulars (S0)& 429                 & 38          & 32                & 70             & 499\\ \hline
     	\textbf{LTG}    & 827                 &  116        & 391               & 507            & 1334\\
                            & (61.99\%)           &  (8.7\%)    & (29.31\%)         & (38.01\%)      & (34.21\%)\\ \hline
 		\textbf{Total}  & 3259                &  193        & 447               & 640            & 3899\\
 		                  & (83.58\%)           &  (4.95\%)   & (11.46\%)         & (16.41\%)      &(100\%)\\
 		                  \hline
 		                  \hline
 		                  \\
 		                  \multicolumn{6}{c}{Field sample} \\
 		                  & \textbf{Passive}    & \textbf{GV} & \textbf{SF}       & \textbf{GV+SF}  & \textbf{Total}\\ \hline
     	\textbf{ETG}    & 4385                &  378        & 715               & 1093            & 5478 \\
 		                  & (80.05\%)           &  (6.90\%)   & (13.05\%)         & (19.95\%)       & (47.80\%)\\
            Ellipticals (E) & 3461                & 188         & 239               & 427             & 3888\\
            Lenticulars (S0)& 924                 & 190         & 476               & 666             & 1590\\ \hline
     	\textbf{LTG}    & 1596                &  827        & 3559              & 4386            & 5982  \\
                            & (26.68\%)           &  (13.82\%)  & (59.50\%)         & (73.32\%)       & (52.20\%)\\ \hline
 		\textbf{Total}  & 4386                &  1205       & 4274              & 4274            & 11460\\
 		                  & (52.19\%)           &  (10.51\%)  & (37.29\%)         & (47.81\%)       &(100\%)\\
 		                  \hline
 		                  \hline
 	\end{tabular}}
    \caption{Morphological division for the galaxy cluster sample (top table) and for the field sample (bottom table). We also show the division between passive, green-valley, and star-forming galaxies (and also a combination of the GV and SF populations). This separation was made using eqs. 1 and 2 (see \S \ref{subsec:star_formation_levels} for more details).}
 	\label{tab:tabela_1}
 \end{table}
%------------------------------------------------------------------------------------------------------------------%

%------------------------------------------------------------------------------------------------------------------%
\begin{figure*}
    \includegraphics[width=\linewidth]{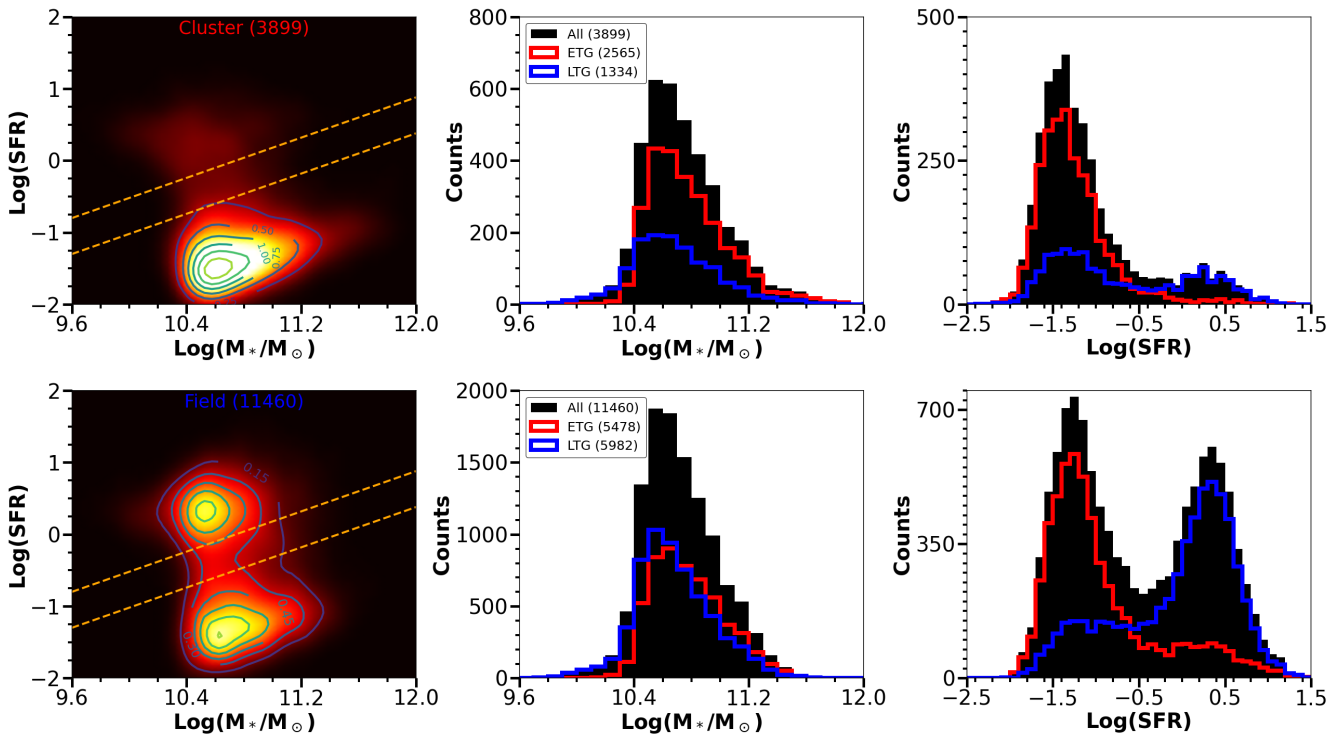}
    \caption{Stellar mass \textit{vs.} SFR diagram. In red, we have the plot for the cluster sample and in blue we have the field sample. In both, the dashed lines are equations (1) and (2) of \citealt{trussler+20} (equations \ref{eq:equation1} and \ref{eq:equation2} here), delineating the passive and the star-forming regions. This figure already gives a good indication of the predominance of passive galaxies in the cluster sample, and the bi-modality observed for the field sample in Table \ref{tab:tabela_1}.}
    \label{fig:sfr_vs_mass}
\end{figure*}

%------------------------------------------------------------------------------------------------------------------%

We combined E and S0 galaxies into one class, which we named Early-type Galaxies (ETGs). Objects that are not ETGs are called Late-type Galaxies (LTGs). When morphological information was not obtainable through DS18, we supplemented the data with the classification provided by HC11. HC11 provided morphological information gathered for the SDSS-DR7 spectroscopic sample (up to $z = 0.25$) via a Machine Learning technique (in their case, using a Support Vector Machine). HC11 gives probabilistic information to classify galaxies as ellipticals, early Spirals (ab), and late Spirals cd. From the $11460$ field galaxies, only $13$ ($\sim 0.1\%$) were absent from the DS18 sample and required HC11 information. From the $3899$ cluster galaxies, only $38$ ($\sim 1\%$) do not have DS18 information and were classified using HC11 data. For these $51$ sources, we utilize the HC11 as follows:

%------------------------------------------------------------------------------------------------------------------%
\begin{itemize}[align=left, itemindent=1em, labelsep=0em, labelwidth=1em, leftmargin=0pt, noitemsep]
    \item \textbf{E:}
    \begin{itemize}[align=left, itemindent=2em, labelsep=0em, labelwidth=1em, leftmargin=0pt, noitemsep]
         \item Prob(E) $> 0.8$
    \end{itemize}
\end{itemize}

\begin{itemize}[align=left, itemindent=1em, labelsep=0em, labelwidth=1em, leftmargin=0pt, noitemsep]
    \item \textbf{S0:}
    \begin{itemize}[align=left, itemindent=2em, labelsep=0em, labelwidth=1em, leftmargin=0pt, noitemsep]
         \item Prob(S0) $> 0.8$
    \end{itemize}
\end{itemize}

We added the morphological classification from HC11 as it is based on a similar data set (also built from the SDSS-DR7 spectroscopic main sample) and it is a classification provided by the same research group as in DS18. Nonetheless, it is important to mention that we consider less than $1\%$ of galaxies from HC11, and removing those galaxies does not affect our main conclusions.
%------------------------------------------------------------------------------------------------------------------%
%------------------------------------------------------------------------------------------------------------------%
\subsection{Separation of passive and star-forming galaxies}
\label{subsec:star_formation_levels}

Galaxy properties such as star formation rate (SFR) and stellar masses were calculated by the Max Planck Institute for Astrophysics and the Johns Hopkins University group (MPA-JHU), following the methods of \citet{Brinchmann+04}, \citet{kau03a}, and \citet{Tremonti+04}. The MPA-JHU sample provides this information for over $1,800,000$ galaxies up to $z \sim 0.3$ for the SDSS-DR8 release. 

Figure \ref{fig:sfr_vs_mass} shows, in the left panels, the M$_*$ \textit{vs.} SFR plane for our cluster sample (top) and field sample (bottom). The dashed lines in each of the panels are the visual construction of the equations (1) and (2) of \cite{trussler+20} (equations \ref{eq:equation1} and \ref{eq:equation2} here). Equation \ref{eq:equation1} gives the boundary between star-forming (SF) and green-valley (GV) galaxies, and Equation \ref{eq:equation2} gives the boundary between GV and passive (Pas) galaxies: 

\begin{equation}
    log (SFR) = 0.70~log (M_*) - 7.52
    \label{eq:equation1}
\end{equation}
\begin{equation}
    log (SFR) = 0.70~log(M_*) - 8.02
    \label{eq:equation2}
\end{equation}

Using the above equations, we separate our data sets (cluster and field galaxies) in different subpopulations, regarding morphology and star-formation activity. Table \ref{tab:tabela_1} summarizes the result of this separation.

To illustrate the impact of environment and morphology on the star formation activity, we also display in Figure \ref{fig:sfr_vs_mass} the stellar mass (central panels) and SFR distributions (right panels) of ETGs (red lines) and LTGs (blue lines). The distributions for all galaxies are shown by the black-filled histograms. As before, the cluster results are in the top panels, while the field results are shown at the bottom. The LTG distributions show fewer massive galaxies (in comparison to the ETGs), both for clusters and the field. However, this difference does not look very large. Despite that, the SFR distributions of LTGs are remarkably different from the ETGs, and those differences are enhanced with the environment (if we compare the cluster and field results). Note the LTG distribution of SFR is very different between the cluster and field datasets. We can also detect an increase of SF ETGs when comparing the cluster and field distributions.
%------------------------------------------------------------------------------------------------------------------%
\subsection{Cluster dynamical state}
\label{subsec: dynamical_state}

To investigate the possible impact of the cluster's dynamical state on the galaxy properties, we also separated the clusters according to their degree of substructures. Objects with no significant substructure are named relaxed, while the others (with strong signs of substructure) are named non-relaxed. To perform this classification, we applied the Dressler $\&$ Shectman test (\citealt{dressler-shctman_88}). For more details, we refer the reader to \citet{pin96, lop06, lop09b, Lopes+18}.

In our cluster sample, within R$_{200}$, we have a total of $2878$ galaxies in $197$ relaxed clusters and $1021$ galaxies in $34$ non-relaxed clusters. A summary of the different galaxy populations found within the relaxed and disturbed systems is given in Table \ref{tab:tabela_2}.
%------------------------------------------------------------------------------------------------------------------%
\begin{table}
\resizebox{\columnwidth}{!}{%
 	\centering
 	\begin{tabular}{lccccc} % six columns, alignment for each
 		                            & \textbf{Passive}    & \textbf{GV}  & \textbf{SF}  &  \textbf{GV+SF}  & \textbf{Total}\\
 		\hline \\
            \multicolumn{6}{c}{Relaxed cluster sample} \\  
 		\textbf{ETG}:               & 1864      & 62       & 45        & 107       & 1971\\
                                        & (94.17\%) & (3.15\%) & (2.28\%)  & (5.57\%)  &(68.49\%) \\
            Ellipticals                 & 1542      & 29       & 21        & 50        & \\
            Lenticulars                 & 322       & 33       & 24        & 57        & \\
            \\
 		\textbf{LTG}:               & 544       & 89       & 274       & 363       & 907\\ 
                                        & (59.98\%) & (9.81\%) & (30.21\%) & (40.02\%) & (31.51\$) \\ 
 		\\\hline
            \multicolumn{6}{c}{Non-relaxed cluster sample} \\  
 		\textbf{ETG}:               & 568       & 15       & 11        & 26        & 594\\ 
                                        & (95.62\%) & (2.53\%) & (1.85\%)  & (4.37\%)  & (58.18\%) \\
            Ellipticals                 & 461       & 10       & 3         & 13        & \\
            Lenticulars                 & 107       & 5        & 8         & 13        & \\
            \\
 		\textbf{LTG}:               & 283       & 27       & 117       & 144       & 427\\ 
                                        & (66.28\%) & (6.32\%) & (27.40\%) & (33.72\%) & (41.48\%)\\ \\ \hline
 		\textbf{Total}              & \textbf{3259} &\textbf{193} & \textbf{447}  & \textbf{640} & 3899\\
 		                            & (83.58\%) & (4.95\%) & (11.46\%) & (16.41\%) & \\ \hline

 	\end{tabular}} 
        \caption{Division for the $3899$ bright galaxies ($M_r \le M^* + 1$) in our cluster sample according to the cluster dynamical state, galaxy morphology, and star-formation activity.}
 	\label{tab:tabela_2}
 \end{table}
%------------------------------------------------------------------------------------------------------------------%

%------------------------------------------------------------------------------------------------------------------%
\section{Results}
\label{sec:results}
%------------------------------------------------------------------------------------------------------------------%
\begin{figure}
	\includegraphics[width=\columnwidth]{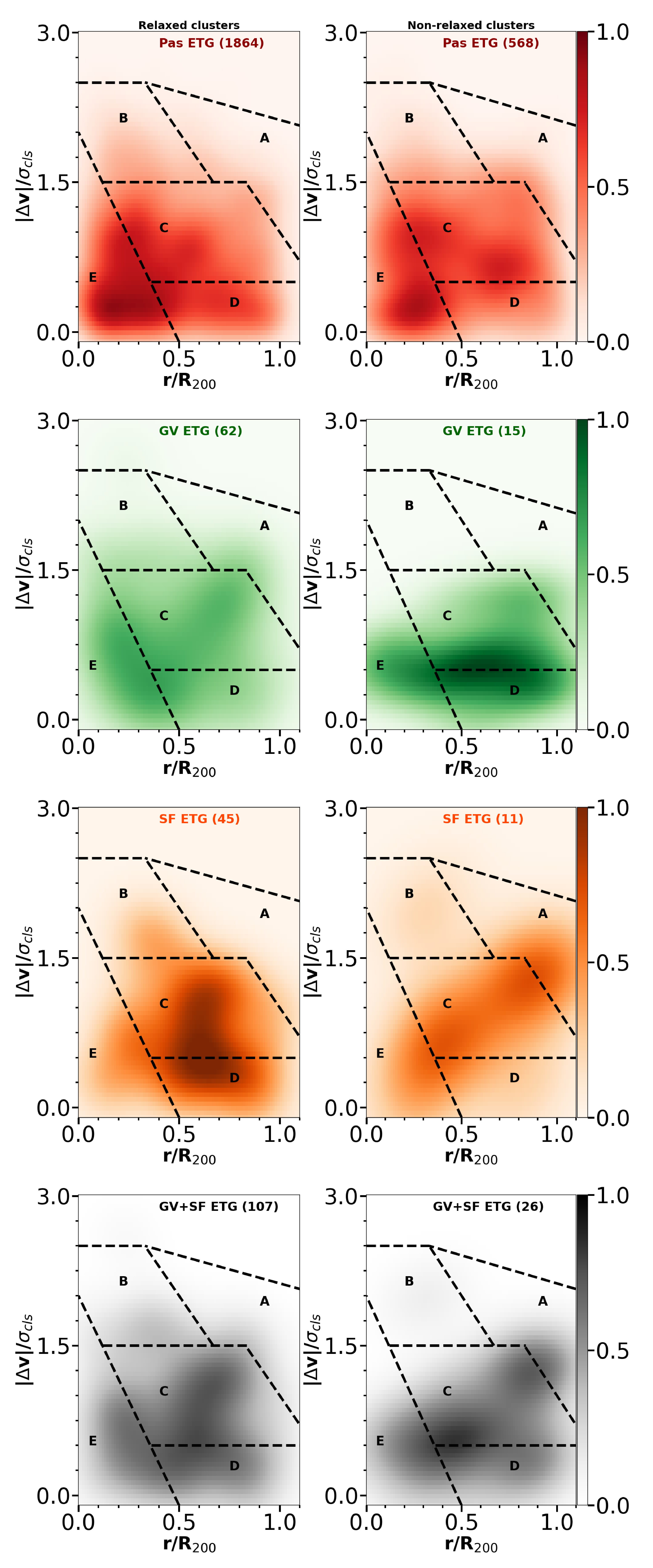}
    \caption{Projected phase-space diagrams for ETGs in relaxed clusters (left side panels) and non-relaxed clusters (right side panels). \citet{rhee+17} regions are delineated by the dashed black lines, and as mentioned in \citet{rhee+17} is an indication of different times since infall. From top to bottom, we show passive (Pas), {\it green valley} (GV), star-forming (SF), and a combination of GV and SF galaxies, respectively.}
    \label{fig:pps_etg}
\end{figure}
%------------------------------------------------------------------------------------------------------------------%
\begin{figure}
	\includegraphics[width=\columnwidth]{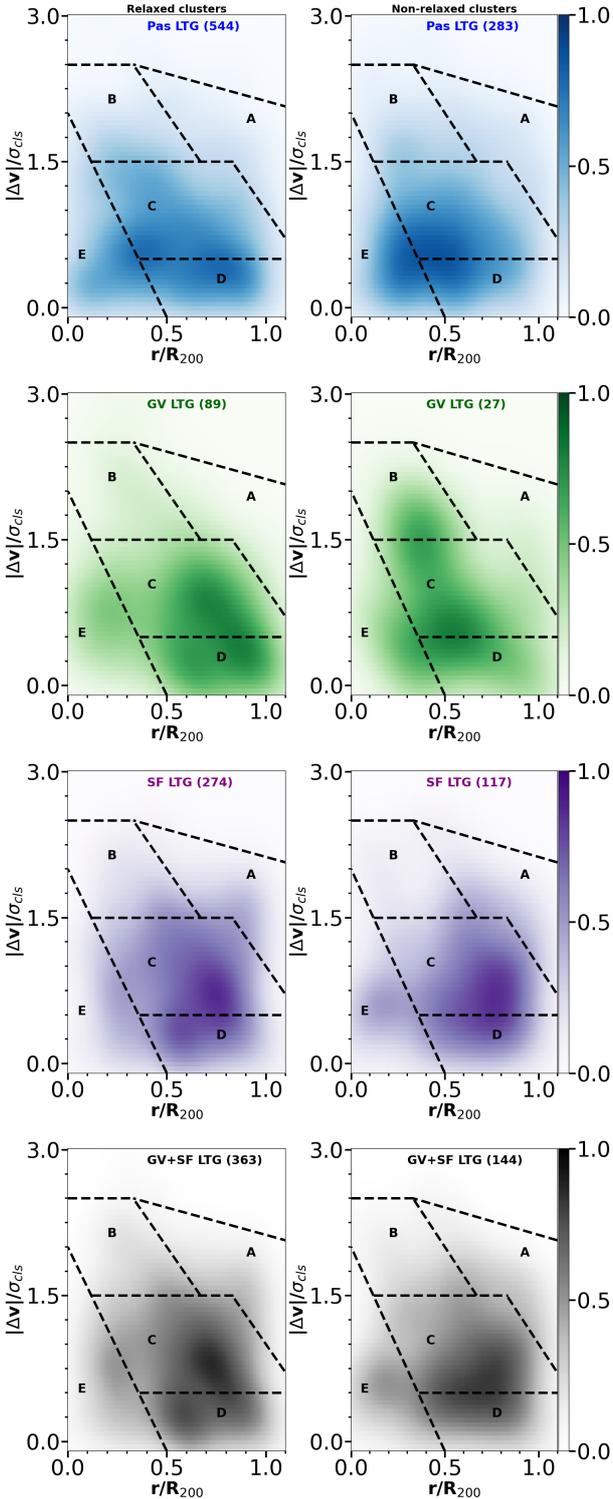}
    \caption{Projected phase-space diagrams for LTGs in relaxed clusters (left side panels) and non-relaxed clusters (right side panels). \citet{rhee+17} regions are delineated by the dashed black lines, and as mentioned in \citet{rhee+17} is an indication of different times since infall. From top to bottom, we show passive (Pas), {\it green valley} (GV), star-forming (SF) and a combination of GV and SF galaxies, respectively.}  
    \label{fig:pps_ltg}
\end{figure}
%------------------------------------------------------------------------------------------------------------------%

The environmental dependence of galaxy populations is clear from Table \ref{tab:tabela_1}. As expected, the ETGs dominate within clusters ($\sim 66\%$ of all galaxies). Different results are found for the field galaxies. LTGs comprise $\sim 52\%$ of all objects in this environment. As for the star-formation activity, we have the following. Inside clusters, the ETGs are overwhelmingly dominated by passive galaxies ($\sim 95\%$). In the field, we found this fraction to be reduced to $\sim 80\%$. For the LTGs, we have a similar trend, but with different values. Within clusters, we found that $\sim 62\%$ of all LTGs are passive, while in the field, this fraction decreases to $\sim 27\%$. This behavior is a strong indication of the environmental influence, despite the morphological classification. In order to have further information about these results, we investigated the location of the galaxies within the projected phase-space (PPS) diagram. We also compare the cumulative distributions of different galaxy properties.
%------------------------------------------------------------------------------------------------------------------%
%------------------------------------------------------------------------------------------------------------------%
\subsection{Location in the Phase-Space Diagram and the impact of the dynamical state}
\label{phase_space+dynamical_state}

As discussed by many authors (e.g., \citealt{Mahajan+11}, \citealt{Oman+13}, \citealt{rhee+17}, \citealt{pasquali+19}, \citealt{pasquali+19}), the projected phase-Space diagram (PPS) can be utilized as a good indicator of the cluster assembly process. Hence, we can use it to estimate how long a galaxy is inside a cluster. Based on a cosmological hydrodynamic simulation, \citet{rhee+17} verified that although the positions of galaxies in the PPS are not as precise as would be the case for the real phase-space diagram, it is still possible and reliable to utilize the PPS as an indicator of the infall time. The authors created a PPS from mock data and divided the diagram into five regions (A to E). These regions are divided to be dominated by four major groups: first infallers (A), recent infallers (B, C), intermediate infallers (D), and ancient infallers (E). Fig. 6 of \cite{rhee+17} gives good visual support to the understanding of these regions. Each one of these categories has a time interval for the time since infall:

%------------------------------------------------------------------------------------------------------------------%
\begin{table*}
\resizebox{\textwidth}{!}{%
\begin{tabular}{ccccccc|}
\multicolumn{7}{l|}{\textbf{Comparison between Relaxed vs. Non-relaxed samples}}                                                    \\ \hline
& \multicolumn{6}{|c|}{\textbf{ETGs}} \\
\multicolumn{1}{c|}{}                                          & 
M$_*$      & 
$\Sigma_5$ & 
D$_n(4000)$   & 
H$\delta$  & 
sSFR       & 
$\nabla$(g - i) \\ \cline{2-7} 
\multicolumn{1}{l|}{Pas$^{relax}$ vs. Pas$^{non-relax}$} &       
\cellcolor[HTML]{9AFF99}$0.1111$ &            
\cellcolor[HTML]{9AFF99}$0.3895$ &          
\cellcolor[HTML]{9AFF99}$0.0773$ &
\cellcolor[HTML]{9AFF99}$0.1412$ &
\cellcolor[HTML]{9AFF99}$0.5596$ &      
\cellcolor[HTML]{FFCCC9}$0.0095$ 
\\
\multicolumn{1}{l|}{GV$^{relax}$ vs. GV$^{non-relax}$} &       
\cellcolor[HTML]{9AFF99}$0.8694$ &            
\cellcolor[HTML]{9AFF99}$0.6954$ &          
\cellcolor[HTML]{9AFF99}$0.8962$ &           
\cellcolor[HTML]{9AFF99}$0.6241$ &
\cellcolor[HTML]{9AFF99}$0.4852$ &      
\cellcolor[HTML]{9AFF99}$0.6167$ 
\\
\multicolumn{1}{l|}{SF$^{relax}$ vs. SF$^{non-relax}$} &       
\cellcolor[HTML]{9AFF99}$0.4086$ &            
\cellcolor[HTML]{9AFF99}$0.5502$ &          
\cellcolor[HTML]{9AFF99}$0.7534$ &           
\cellcolor[HTML]{9AFF99}$0.3111$ &
\cellcolor[HTML]{9AFF99}$0.3588$ &      
\cellcolor[HTML]{9AFF99}$0.8341$ 
\\
\multicolumn{1}{l|}{GV$+$SF$^{relax} vs. GV$+$SF^{non-relax}$} &       
\cellcolor[HTML]{9AFF99}$0.7525$ &            
\cellcolor[HTML]{9AFF99}$0.6059$ &          
\cellcolor[HTML]{9AFF99}$0.7130$ &           
\cellcolor[HTML]{FFCCC9}$3.539e-5$ &
\cellcolor[HTML]{9AFF99}$0.8588$ &      
\cellcolor[HTML]{9AFF99}$0.8218$                  \\ \hline
%-----------------------------------------------------------------------------------------------
& \multicolumn{6}{|c|}{\textbf{LTGs}} \\
\multicolumn{1}{c|}{}                                          & 
M$_*$      & 
$\Sigma_5$ & 
D$_n(4000)$   & 
H$\delta$  & 
sSFR       & 
$\nabla$(g - i) \\ \cline{2-7} 
\multicolumn{1}{l|}{Pas$^{relax}$ vs. Pas$^{non-relax}$} &       
\cellcolor[HTML]{9AFF99}$0.4322$ &      
\cellcolor[HTML]{9AFF99}$0.6392$ &      
\cellcolor[HTML]{9AFF99}$0.7693$ &      
\cellcolor[HTML]{9AFF99}$0.081$ &      
\cellcolor[HTML]{9AFF99}$0.5906$ &      
\cellcolor[HTML]{FFCCC9}$0.0343$       
\\
\multicolumn{1}{l|}{GV$^{relax}$ vs. GV$^{non-relax}$} &       
\cellcolor[HTML]{FFCCC9}$0.0415$ &      
\cellcolor[HTML]{9AFF99}$0.6340$ &      
\cellcolor[HTML]{9AFF99}$0.5264$ &      
\cellcolor[HTML]{9AFF99}$0.7919$ &      
\cellcolor[HTML]{9AFF99}$0.3132$ &      
\cellcolor[HTML]{9AFF99}$0.4825$       
\\
\multicolumn{1}{l|}{SF$^{relax}$ vs. SF$^{non-relax}$} &       
\cellcolor[HTML]{9AFF99}$0.2021$ &      
\cellcolor[HTML]{FFCCC9}$0.0446$ &      
\cellcolor[HTML]{9AFF99}$0.9502$ &      
\cellcolor[HTML]{9AFF99}$0.5258$ &      
\cellcolor[HTML]{9AFF99}$0.9744$ &      
\cellcolor[HTML]{9AFF99}$0.1652$ 
\\
\multicolumn{1}{l|}{GV$+$SF$^{relax} vs. GV$+$SF^{non-relax}$} &       
\cellcolor[HTML]{FFCCC9}$8.839e-3$ &      
\cellcolor[HTML]{9AFF99}$0.080$ &      
\cellcolor[HTML]{9AFF99}$0.3160$ &      
\cellcolor[HTML]{FFCCC9}$6.625e-5$ &      
\cellcolor[HTML]{9AFF99}$0.4629$ &      
\cellcolor[HTML]{9AFF99}$0.1711$                  \\ \hline \hline \\
%-----------------------------------------------------------------------------------------------
%-----------------------------------------------------------------------------------------------
\multicolumn{7}{l|}{\textbf{Comparison between Cluster vs. Field samples}} \\ \hline
& \multicolumn{6}{|c|}{\textbf{ETGs}} \\
\multicolumn{1}{c|}{}                                          & 
M$_*$      & 
$\Sigma_5$ & 
D$_n(4000)$   & 
H$\delta$  & 
sSFR       & 
$\nabla$(g - i) \\ \cline{2-7} 
\multicolumn{1}{l|}{Pas$^{cluster}$ vs. Pas$^{field}$} &       
\cellcolor[HTML]{FFCCC9}$2.978e-6$ &            
\cellcolor[HTML]{FFCCC9}$1.163e-10$ &            
\cellcolor[HTML]{FFCCC9}$6.661e-15$ &            
\cellcolor[HTML]{FFCCC9}$1.110e-16$ &            
\cellcolor[HTML]{FFCCC9}$1.110e-16$ &            
\cellcolor[HTML]{FFCCC9}$4.219e-15$            
\\
\multicolumn{1}{l|}{GV$^{cluster}$ vs. GV$^{field}$} &       
\cellcolor[HTML]{9AFF99}$0.078$ &            
\cellcolor[HTML]{FFCCC9}$0.000$ &            
\cellcolor[HTML]{9AFF99}$0.4692$ &            
\cellcolor[HTML]{FFCCC9}$1.331e-4$ &            
\cellcolor[HTML]{9AFF99}$0.9039$ &            
\cellcolor[HTML]{FFCCC9}$8.523-8$             
\\
\multicolumn{1}{l|}{SF$^{cluster}$ vs. SF$^{field}$} &       
\cellcolor[HTML]{9AFF99}$0.2319$ &            
\cellcolor[HTML]{FFCCC9}$1.110e-16$ &            
\cellcolor[HTML]{9AFF99}$0.0554$ &            
\cellcolor[HTML]{FFCCC9}$0.0159$ &            
\cellcolor[HTML]{9AFF99}$0.073$ &            
\cellcolor[HTML]{FFCCC9}$0.0132$             
\\
\multicolumn{1}{l|}{GV$+$SF$^{cluster} vs. GV$+$SF^{field}$} &       
\cellcolor[HTML]{9AFF99}$0.087$ &            
\cellcolor[HTML]{FFCCC9}$3.331e-16$ &            
\cellcolor[HTML]{FFCCC9}$4.806e-6$ &            
\cellcolor[HTML]{FFCCC9}$1.489e-20$ &            
\cellcolor[HTML]{FFCCC9}$1.521e-7$ &            
\cellcolor[HTML]{FFCCC9}$4.922e-8$             
\\ \hline
%-----------------------------------------------------------------------------------------------
& \multicolumn{6}{|c|}{\textbf{LTGs}} \\
\multicolumn{1}{c|}{}                                          & 
M$_*$      & 
$\Sigma_5$ & 
D$_n(4000)$   & 
H$\delta$  & 
sSFR       & 
$\nabla$(g - i) \\ \cline{2-7} 
\multicolumn{1}{l|}{Pas$^{cluster}$ vs. Pas$^{field}$} &       
\cellcolor[HTML]{FFCCC9}$1.228e-6$ &            
\cellcolor[HTML]{FFCCC9}$2.109e-15$ &            
\cellcolor[HTML]{FFCCC9}$1.631e-6$ &            
\cellcolor[HTML]{FFCCC9}$2.109e-15$ &            
\cellcolor[HTML]{FFCCC9}$2.109e-15$ &            
\cellcolor[HTML]{FFCCC9}$5.815e-38$             
\\
\multicolumn{1}{l|}{GV$^{cluster}$ vs. GV$^{field}$} &       
\cellcolor[HTML]{FFCCC9}$7.741e-5$ &            
\cellcolor[HTML]{FFCCC9}$9.992e-16$ &            
\cellcolor[HTML]{FFCCC9}$0.012$ &            
\cellcolor[HTML]{9AFF99}$0.6756$ &            
\cellcolor[HTML]{FFCCC9}$0.014$ &            
\cellcolor[HTML]{FFCCC9}$2.781e-7$             
\\
\multicolumn{1}{l|}{SF$^{cluster}$ vs. SF$^{field}$} &       
\cellcolor[HTML]{FFCCC9}$5.551e-16$ &            
\cellcolor[HTML]{FFCCC9}$5.551e-16$ &            
\cellcolor[HTML]{FFCCC9}$5.145e-6$ &            
\cellcolor[HTML]{FFCCC9}$0.016$ &            
\cellcolor[HTML]{FFCCC9}$8.978e-5$ &            
\cellcolor[HTML]{FFCCC9}$7.923e-10$             
\\
\multicolumn{1}{l|}{GV$+$SF$^{cluster} vs. GV$+$SF^{field}$} &       
\cellcolor[HTML]{FFCCC9}$2.442e-15$ &            
\cellcolor[HTML]{FFCCC9}$3.331e-16$ &            
\cellcolor[HTML]{FFCCC9}$7.488e-4$ &            
\cellcolor[HTML]{FFCCC9}$1.196e-6$ &            
\cellcolor[HTML]{FFCCC9}$0.003$ &            
\cellcolor[HTML]{FFCCC9}$4.108e-15$             
\\ \hline
\end{tabular}
}
\caption{P-values for the KS tests comparing our different sub-samples. The first batch of tests compares galaxies inside relaxed clusters with galaxies inside non-relaxed clusters (comparing stellar mass, the local galaxy density ($\Sigma_5$), $4000$ {\AA} break, H$\delta$, specific star formation rate, and color gradient). The second batch of tests compares galaxies in the combined cluster sample (relaxed+non-relaxed clusters) with galaxies in the field. Consider that  $\leq 0.05$ the null hypothesis is refuted \textbf{with $2\sigma$ significance }(distinguishable sample $-$ red background), and for p-value, $ > 0.05$ the null hypothesis is accepted \textbf{with $2\sigma$ significance} (indistinguishable sample $-$ green background).}
\label{tab:tabela_3}
\end{table*}
%------------------------------------------------------------------------------------------------------------------%

%------------------------------------------------------------------------------------------------------------------%
\begin{itemize}[align=left, itemindent=1em, labelsep=0em, labelwidth=1em, leftmargin=0pt, noitemsep]
    \item First Infallers (A): Not fallen yet
    \item Recent Infallers (B and C): $0.00 <  t_{inf} < 3.63 Gyr$
    \item Intermediate Infallers (D): $~3.63 < t_{inf}  < 6.45 Gyr$
    \item Ancient Infallers (E): $~~~~~~~~~6.45 < t_{inf}  < 13.7 Gyr$
\end{itemize}
%------------------------------------------------------------------------------------------------------------------%

In Figures \ref{fig:pps_etg} and \ref{fig:pps_ltg}, we show the PPS of our data, for ETGs and LTGs, respectively. The data is divided into relaxed (left side panels) and non-relaxed clusters (right side panels), and also subdivided into the subpopulations present for each morphology: Pas (first row), GV (second row), and SF (third row). We further added a combination of GV and SF galaxies (fourth row). The dashed lines in each diagram delimit the regions introduced by \citet{rhee+17}.

Figure \ref{fig:pps_etg} shows the differences in the distributions of ETGs in the PPS as a function of their star-formation activity. Pas-ETGs have a higher concentration in the E region (dominated by galaxies in the ancient infall time interval). On the contrary, the GV-ETGs and SF-ETGs tend to avoid this region. These galaxies are predominantly found in the regions labeled C and D (dominated by galaxies in the recent/intermediate infall time interval). The tendency is also present when we compare the Pas-ETGs with the combined sample of GV$+$SF-ETGs. 

The distinction present in Figure \ref{fig:pps_etg} indicates that Pas and GV and/or SF-ETGs inhabit different regions of the PPS, thus having distinct times since infall: shorter periods for GV/SF-ETGs in comparison to the Pas-ETGs counterparts. As those galaxies have the same morphology but different infall times and different levels of star-formation activity, we can formulate some hypotheses: (i) the GV/SF-ETGs enter the clusters as such, and we are witnessing their residual star-formation; (ii) the GV/SF-ETGs are going through a new star-formation event, bringing them back from the passive phase $-$ rejuvenation process that could be triggered by gas infusion \citep{yi+05, marino+09, kaviraj+09, kaviraj+11}.

%------------------------------------------------------------------------------------------------------------------%
\begin{figure*}
\includegraphics[width=2.1\columnwidth]{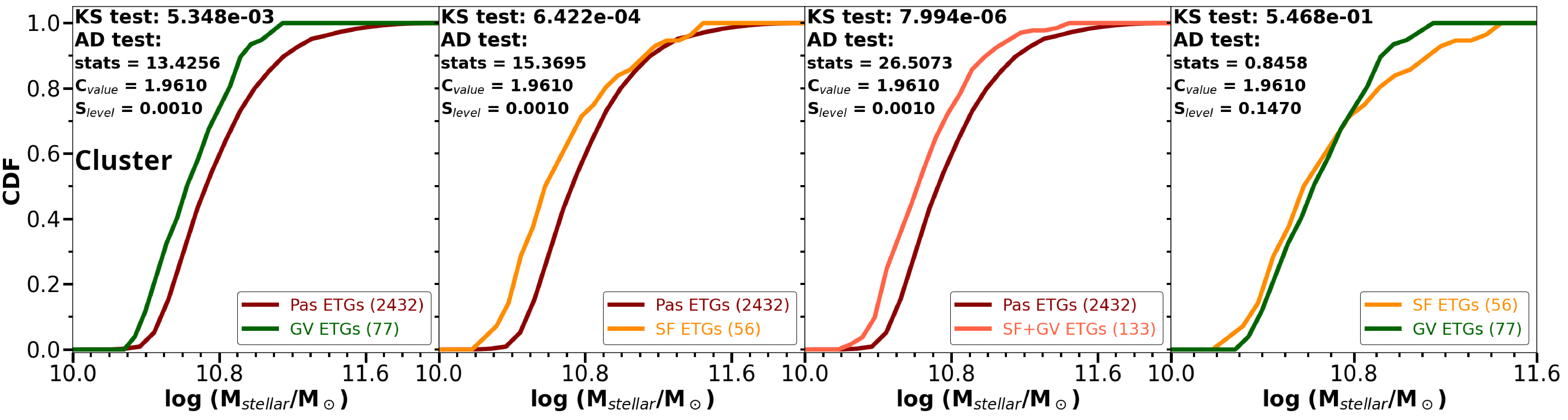}
    \caption{Cumulative Distribution Functions of the local galaxy density ($\Sigma_5$) comparing the cluster subpopulations of ETGs and LTGs. In the first row of panels, the cluster ETGs are the ones being compared. In this instance, red is Pas-ETGs, green is GV-ETGs, orange is SF-ETGs and salmon is GV+SF-ETGs. In the second row of panels, cluster LTGs are the ones being compared. In this instance, in blue are the Pas-LTGs, in green are the GV-LTGs, in purple are the SF-LTGs, and finally, in salmon the combined GV+SF-LTGs. Since the p-value for the local density parameter in Table \ref{tab:tabela_1} shows that the distribution of log ($\Sigma_5$) is drawn from the same distribution for all subpopulations in the relaxed and non-relaxed samples (with an exception for SF-LTGs), the present plot was made by combining the subpopulation from the relaxed sample with their counterpart in the non-relaxed sample. The numbers in the top part of the CDF plots are the p-values for those properties (see Table \ref{tab:tabela_3}).}
    \label{fig:cdf_sigma_5}
\end{figure*}
%------------------------------------------------------------------------------------------------------------------%

Another aspect worth noticing here is the apparent absence of similar behavior for the LTGs in Figure \ref{fig:pps_ltg}. There is a large spread in the distribution of LTGs in the PPS for the three subpopulations (Pas, GV, and SF). The regions they occupy are similar (mainly C and D), pointing to similar infall times. It is also interesting to note that these regions are similar to those occupied by the GV and SF-ETGs. That could indicate the GV and SF ETGs and LTGs could be subject to similar environmental effects, despite their morphology. On the other hand, the similar location of the passive and GV/SF-LTGs may indicate that the former did not have time yet to go through a morphological transformation but may have already quenched their star formation. Note that although the GV and SF-LTG subpopulations occupy similar regions in the PPS to the Pas-LTGs, they do have small distinctions. For instance, we find different fractions of Pas/GV/SF-LTGs in region E. In the case of relaxed clusters, the Pas-LTGs have $35.85\%~\pm~2.06\%$ of its galaxies inside region E, while the SF-LTGs have $24.82\%~\pm~2.61\%$. For the non-relaxed clusters, the scenario is similar, Pas-LTGs have $33.92\%~\pm~2.81\%$, and the SF-LTGs have $23.08\%~\pm~3.90\%$.

Besides the location in the PPS in Figures \ref{fig:pps_etg} and \ref{fig:pps_ltg}, in Figure \ref{fig:cdf_sigma_5} we compare the local density parameter ($\Sigma_5$). We do so to investigate if the cluster action in the GV and SF-ETGs is a more local influence than a global one. For ETGs in clusters, the passive galaxies inhabit the same environment as the GV counterpart, while being statistically distinct from the SF counterpart. But, the GV and the SF seem to inhabit the same environment. Pas-LTGs occupy regions with higher local densities than SF-LTGs and GV+SF-LTGs. For the Pas-LTGs and GV-LTGs, the distinctions are not so strong, with the KS and AD tests very close to the confidence level. That could indicate that these passive objects may have been inside the clusters for slightly longer periods compared to the GV and SF galaxies, especially for ETGs. Another possibility is that these passive galaxies have been already pre-processed in less massive groups before infalling into the clusters \citep{Jaffe+12, Jaffe+15, haines+15, mahajan+13}.

Despite that, for the cluster sample, the results point towards a transition of environments where the GV galaxies are in a local environment very similar to the Pas ones (but not quite identical since the KS test and AD test give values close to the confirmation of the null hypothesis), and the SF galaxies are located in a less dense environment than both of them. Figures \ref{fig:pps_etg} and \ref{fig:pps_ltg} already hint in this direction by showing that Pas-ETGs preferentially inhabit the central regions of the cluster more than their SF counterparts, where the local density is higher than the outer parts of the cluster.  
    
%------------------------------------------------------------------------------------------------------------------%
\begin{figure*}
    \includegraphics[width=2.1\columnwidth]{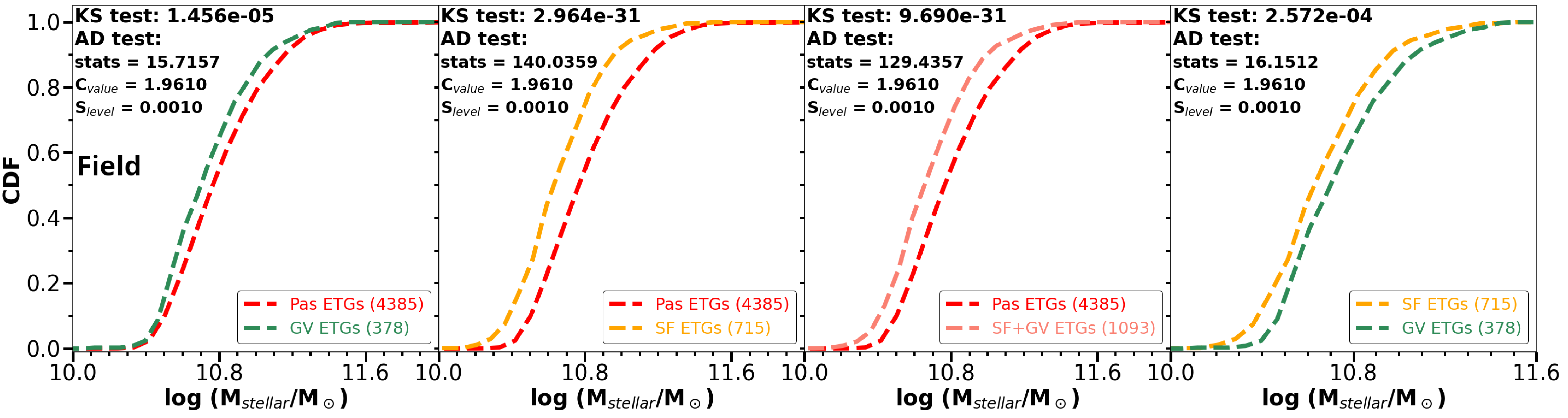}
    \includegraphics[width=2.1\columnwidth]{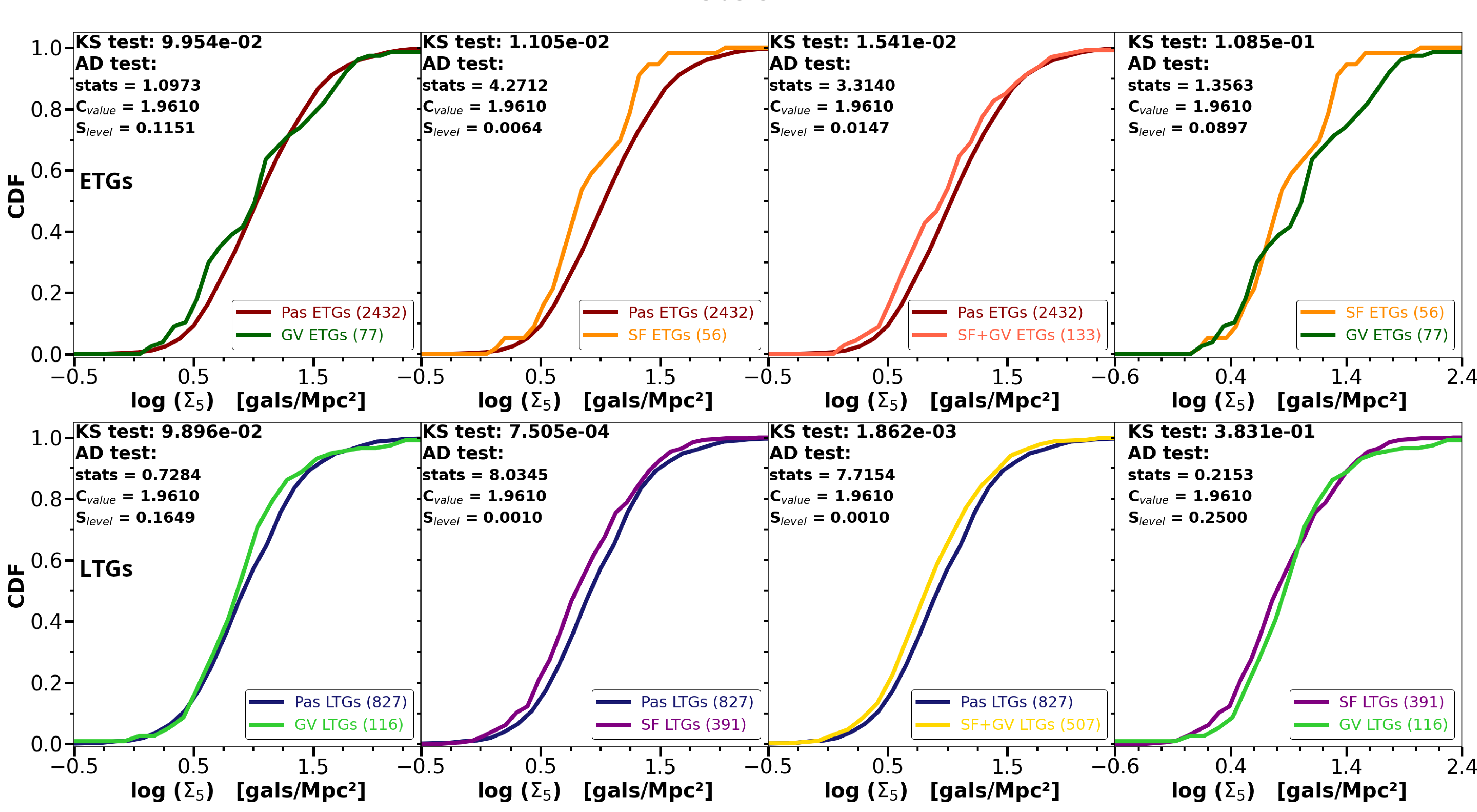}
		
    \caption{Cumulative distribution function for stellar masses. The top panels compare the cluster ETGs (continuous lines), while the bottom panels compare the field ETGs (dashed lines). In both cases, red tones correspond to Pas-ETGs, green tones are GV-ETGs, orange tones are SF-ETGs, and salmon tones the combine GV+SF ETGs. The numbers in the top part of the CDF plots are the p-values for those properties (see Table \ref{tab:tabela_3}).}
    \label{fig:cdf_stellar_mass}
\end{figure*}
%------------------------------------------------------------------------------------------------------------------%

Having established that the infall time of the passive galaxies (even more for Pas-ETGs, but also true for Pas-LTGs) is different from the other populations and that there may be small differences among the latter, we then proceed to investigate the distributions of different galaxies and environmental properties. To do so, we compare different properties by applying the Kolmogorov-Smirnov (KS) test to our sample (Table \ref{tab:tabela_3}). The $p-value$ obtained by comparing the different subpopulations in the relaxed and non-relaxed samples (displayed in Table \ref{tab:tabela_3}) indicates that the dynamical state of the cluster bears no significant influence on the properties studied in this work (with few exceptions that can be observed in the Table). This result is also supported by Table \ref{tab:tabela_2}. Although Table \ref{tab:tabela_2} shows a small discrepancy in the GV and SF fractions of galaxies in relaxed and non-relaxed clusters, it is not a major difference, even more, because the number of galaxies in the non-relaxed clusters is much smaller than available in the relaxed sample. The indication provided by the p-values obtained with the KS test points to a scenario where, if the cluster environment influences the galaxy evolution, the conditions for this influence exist independent of the cluster dynamical state. 

In order to give more confidence to this result, we also made the same comparison shown in Table \ref{tab:tabela_3} using the Anderson-Darling test. The Anderson-Darling test is a variation of the KS test that gives more weight to the tails of the distribution than the KS test (more centrally focused). The results were almost identical, at a $5\%$ significance level, to the ones obtained from the KS test. For the ETGs, the only divergence was for the D$_n(4000)$ when comparing Pas-ETGs in the relaxed and non-relaxed cluster where the KS test gives a p-value for the samples being drawn from the same distribution, but the AD test does not. For the LTGs, the discrepancy is present in three cases. First, in the comparison for stellar mass between GV-LTGs in relaxed and non-relaxed samples, the KS test rejects the null hypothesis, but the AD test accepts it. The second discrepancy for LTGs is for the local density for the comparison between GV+SF-LTGs in relaxed and non-relaxed, where the KS test accepts the null hypothesis, but the AD test does not. The third discrepancy in LTGs is for the color gradient in the comparison between Pas-LTGs in relaxed and non-relaxed samples, where the KS rejects the null hypothesis and the AD test accepts.
 
Figure \ref{fig:cdf_stellar_mass} shows the distribution of stellar mass for the cluster sample in the top panels (considering different subpopulations of ETGs). Since the KS test indicates that the samples are statistically similar, we combine both relaxed and non-relaxed samples to compare the mass distribution of each one of the subpopulations. For the cluster sample, we can see that the GV and SF galaxies (also GV+SF galaxies) are systematically less massive than their passive counterparts.

Two caveats should be considered when looking at the comparisons in Figures \ref{fig:pps_etg} to \ref{fig:cdf_stellar_mass}. The first is the large time intervals for each of the \citet{rhee+17} regions. Naturally, a finer tune in those regions would give a better grasp of the galaxies' infall intervals and their preferred locations in the PPS as a function of morphology and star-formation activity. The second is that when we compare galaxies in  clusters with different dynamical states, we have to be careful since in a few cases we have just a few objects (especially for the GV and SF-ETGs in non-relaxed clusters), leading to small statistic confidence. In this sense, it would also be possible that we are only ruling out only large distinctions, not the subtle ones. Taking these caveats into account, it seems that the positioning of different galaxy populations in the PPS could point to small differences between the relaxed and non-relaxed clusters, but a larger sample is needed in order to reach definitive conclusions. 

To further investigate the influence of the cluster environment in the galaxy evolution, in the next section, we compare the cluster and field samples.

%------------------------------------------------------------------------------------------------------------------%
%------------------------------------------------------------------------------------------------------------------%
\begin{figure*}
	\includegraphics[width=\linewidth]{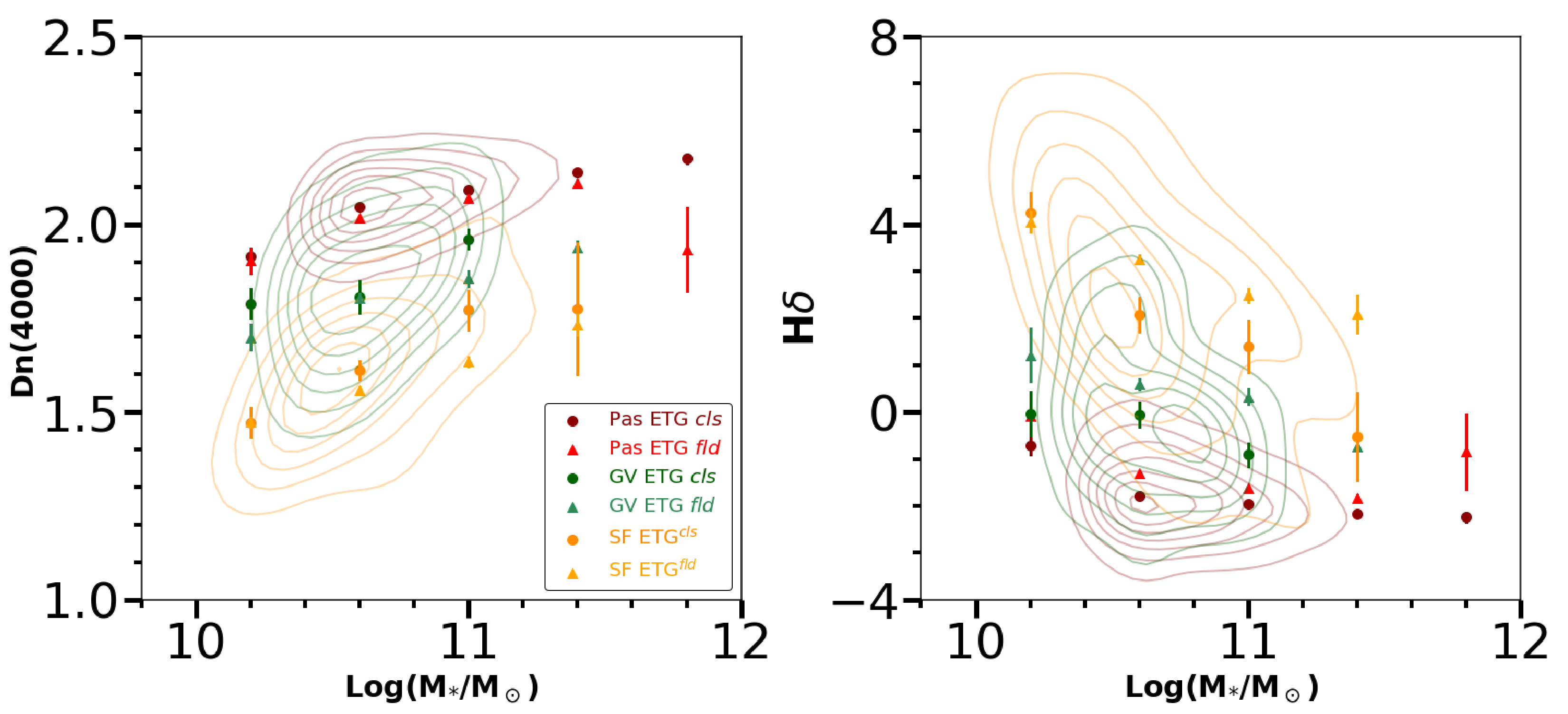}
	\includegraphics[width=\linewidth]{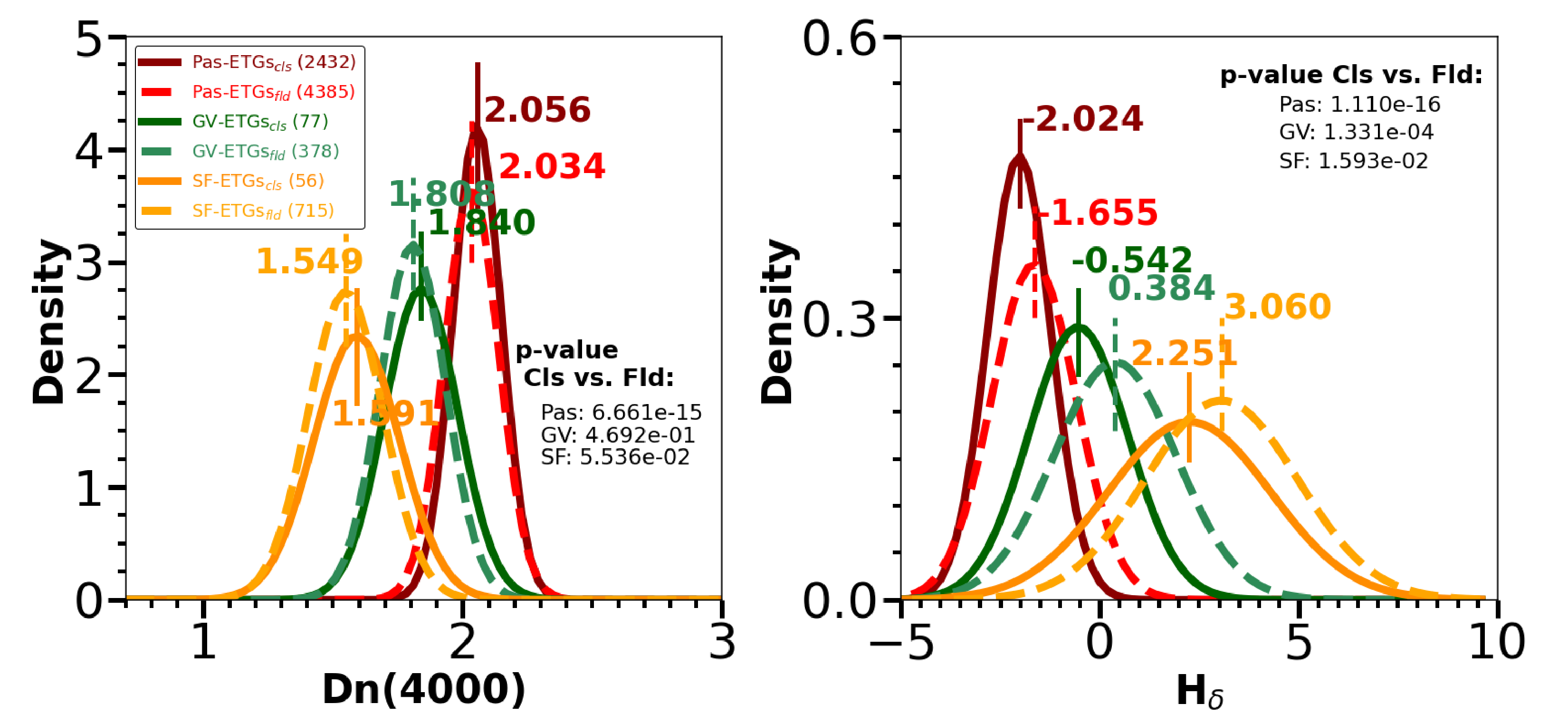}
    \caption{D$_n(4000)$ \textit{vs.} M$_*$ (top left panel) and $H\delta$ \textit{vs.}  M$_*$ (top right panel) planes. The contour curves illustrate the cluster distribution. The dots are the median value for the binned mass for clusters, while the triangles are the median value for the binned mass for the field sample. As can be seen, for fixed stellar mass, the median values of D$_n(4000)$ (H$\delta$) for GV and SF galaxies are always lower (bigger) than for Pas. The same behavior for the SF in comparison to GV is observed. In what regards the cluster \textit{vs.} field comparison, small distinctions are observed in this plane. In the bottom panels, we show the Gaussian fit for each of the distributions. The distribution shows clearly a distinction between the Pas, GV, and SF-ETGs. On the other hand, when comparing cluster \textit{vs.} field, for the D$_n(4000)$ the distinction is nonexistent for GV and SF-ETGs (but present for Pas-ETGs). While, for H$\delta$, they are all distinct from each other. This result is given by the p-value of the KS test.}
    \label{fig:cdf_dn4000_hdelta}
\end{figure*}
%------------------------------------------------------------------------------------------------------------------%

\subsection{The impact of the global environment -- cluster \textit{vs.} field}
\label{cluster_vs_field}

%------------------------------------------------------------------------------------------------------------------%
\begin{figure*}
    \includegraphics[width=\linewidth]{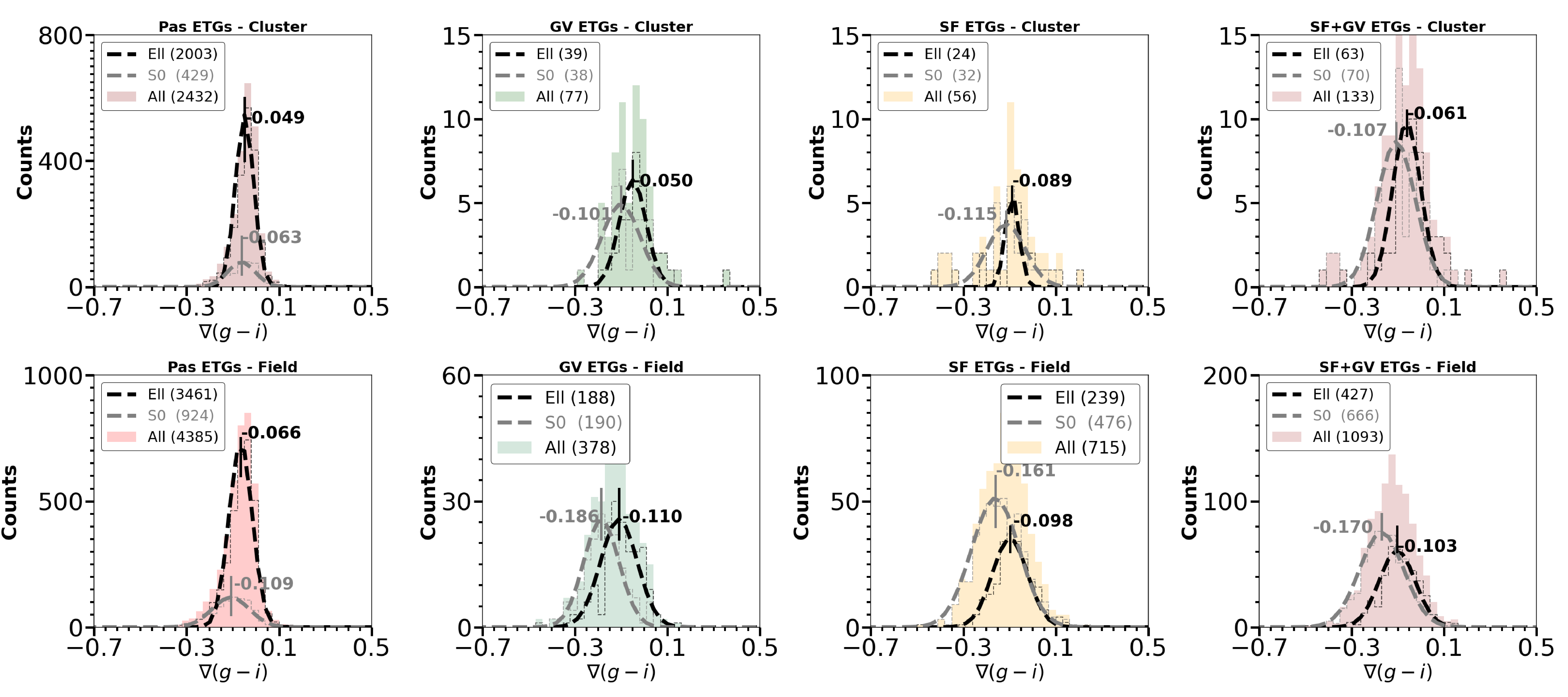}
    \caption{Distribution of $\nabla(g - i)$ for each of the subpopulations in the cluster (continuous lines) and field (dashed lines) samples. Red tones represent Pas-ETGs, green tones GV-ETGs, and orange tones SF-ETGs. The numbers in the legend are the mean value of the Gaussian fit, while the top right numbers are the p-values for the comparison between the cluster and field subpopulations. As can be easily seen, the color gradient distributions are different when we compare Pas, GV, and SF ETGs between themselves, and also when we compare the cluster populations with their counterparts in the field.}
    \label{fig:color_grad_distribution}
\end{figure*}

%------------------------------------------------------------------------------------------------------------------%
%------------------------------------------------------------------------------------------------------------------%

%------------------------------------------------------------------------------------------------------------------%
\begin{figure*}
    \includegraphics[width=\linewidth]{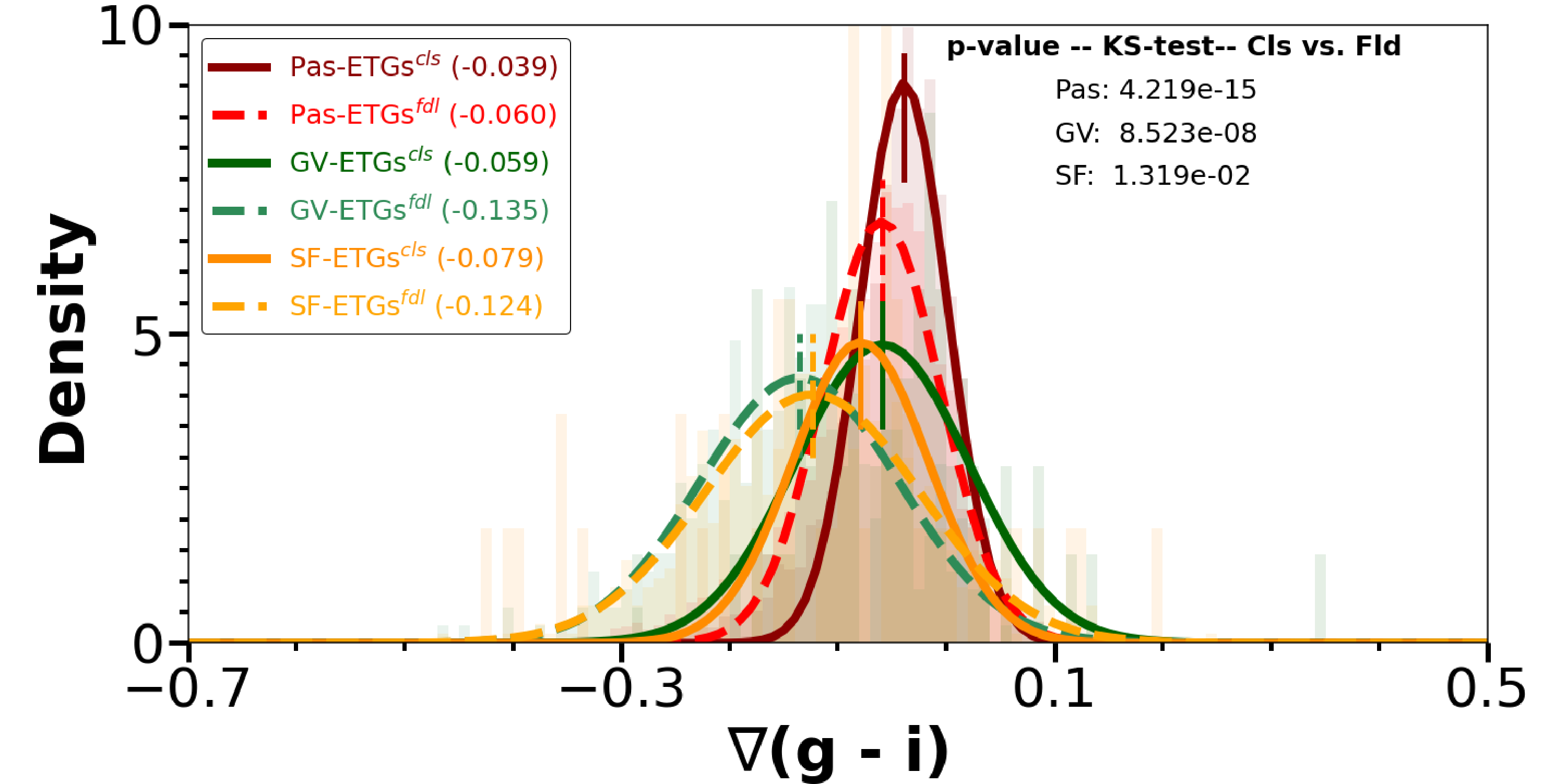}
    \caption{Distributions of the color gradient for the ETGs, split into elliptical (thin black dashed lines) and S0 (thin gray dashed lines) galaxies. The colored histograms represent the distribution for the whole sample. The top panels present the distribution for cluster ETGs, and the bottom panels present the distribution for field ETGs. The vertical lines represent the median values (listed in all panels) of the Gaussian fit for each distribution.}
    \label{fig:hist_color_grad}
\end{figure*}
%------------------------------------------------------------------------------------------------------------------%

On the contrary direction from the comparison between the relaxed and non-relaxed environments, the comparison between cluster and field samples bears significant differences (also Table \ref{tab:tabela_3}. When we compare the galaxies in the field with those in the combined cluster sample (a combination of the relaxed and non-relaxed samples), the great majority of the p-values for the KS test have values below $0.05$ (displayed in the bottom of Table \ref{tab:tabela_3}). There are some cases where the p-value is $>0.05$, and it is important to highlight them $-$ the stellar mass for GV, SF, and GV+SF-ETGs accept the null hypothesis, the same occurs for the D$_n(4000)$ and sSFR for the GV and SF-ETGs comparisons. Overall, field and cluster samples are distinguishable between themselves. We applied the Anderson-Darling test to these samples to confirm the results obtained by the KS test. At a $5\%$ significance level, the great majority of the KS tests are confirmed by the AD test, with four discrepancies for the ETGs and one for the LTGs. The discrepancies for the ETGs occur for M$_*$, both for the GV-ETGs and the GV+SF-ETGs, as well as for D$_n(4000)$ and sSFR in the case of the SF-ETGs. For the LTGs, the discrepancy occurs for the sSFR of GV galaxies. While the KS test refuses the null hypothesis, the AD test accepts it. Besides the summary displayed in Table \ref{tab:tabela_3}, we also show the comparison of some properties between the field and cluster populations in Figures \ref{fig:cdf_dn4000_hdelta} and \ref{fig:color_grad_distribution}.
%------------------------------------------------------------------------------------------------------------------%
\begin{figure*}
    \includegraphics[width=\linewidth]{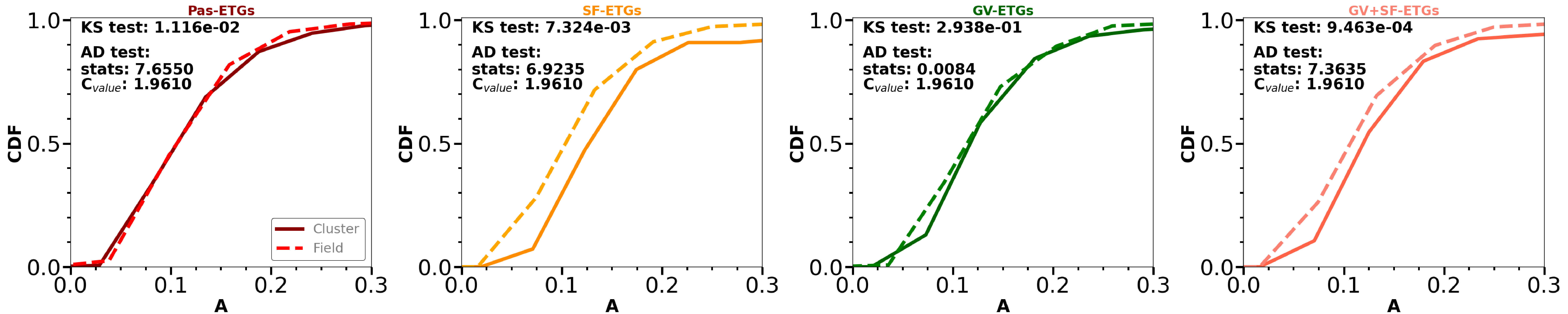}
    \includegraphics[width=\linewidth]{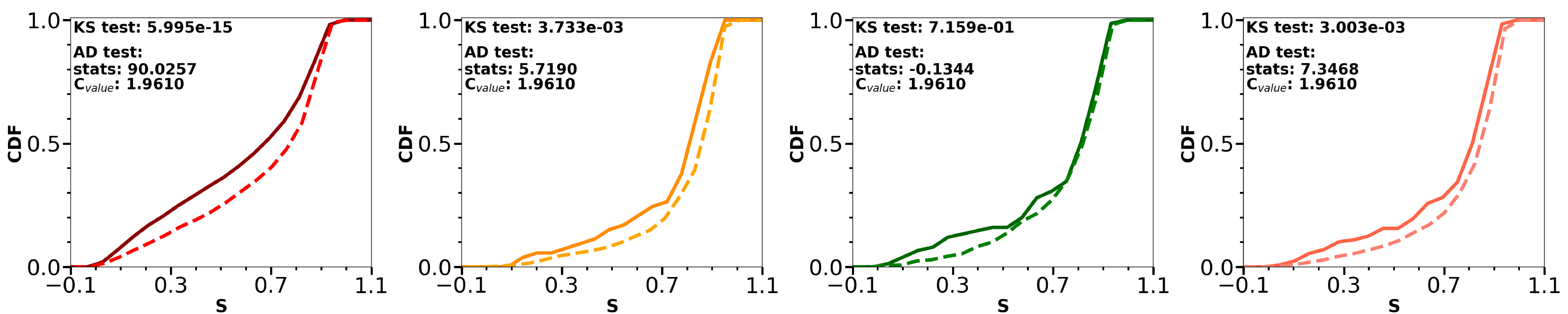}
    \caption{Comparison of CDFs for Asymmetry and Smoothness for all subpopulations. As in previous figures, continuous lines represent cluster samples, and dashed lines represent field galaxies. Pas-ETGs are in red (first column), SF-ETGs in orange (second column), GV-ETGs in green (third column), and GV+SF-ETGs in salmon (fourth column). The top numbers show the p-value of the KS test and the statistic and critical value for the $95\%$ confidence level for the Anderson-Darling test (the null hypothesis is accepted when $stats~<~C_{value}$).}
    \label{fig:cdf_A_and_S}
\end{figure*}
%------------------------------------------------------------------------------------------------------------------%

%------------------------------------------------------------------------------------------------------------------%
\begin{figure}
    \centering
    \includegraphics[width=0.48\textwidth]{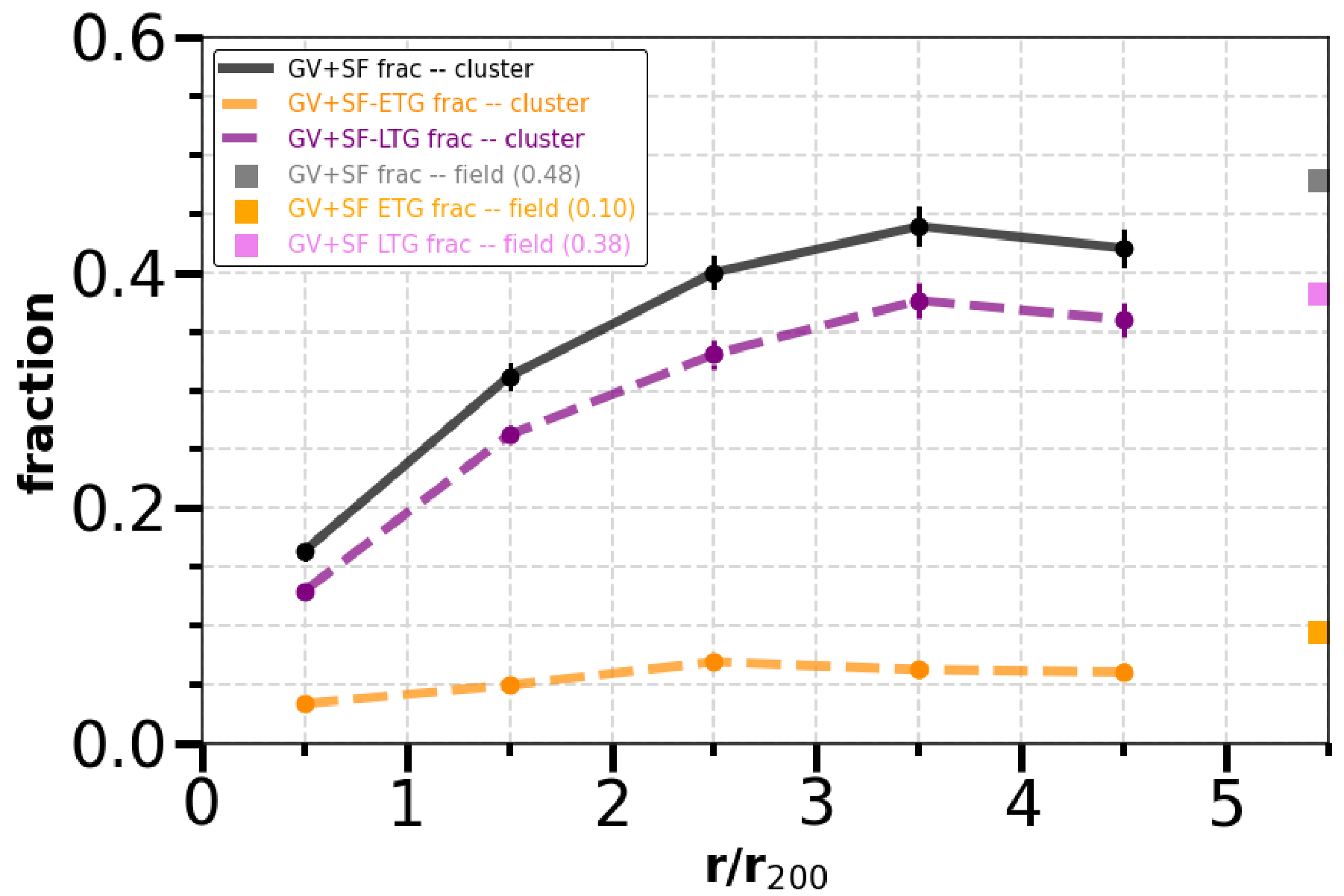}
    \caption{Star-forming fractions variations with clustercentric distance (normalized by R$_{200}$). The continuous black line shows the variation of the overall fractions of star-forming galaxies without morphological distinction. In this figure, we are considering as SF galaxies the combination of GV and SF galaxies. The orange dashed curve only considers SF-ETGs, and the purple dashed line displays the results for SF-LTGs. The gray square in the extreme right part of the figure is the fraction for all star-forming galaxies in the field. We have also indicated the field values for SF-ETGs and SF-LTGs, with the orange and purple squares, respectively. The error bars indicate the standard error for proportions.}
    \label{fig:sf_frac}
\end{figure}
%------------------------------------------------------------------------------------------------------------------%

Of the properties shown in Table \ref{tab:tabela_3}, two of them give us a clue about the star-formation history of the galaxies. They are the D$_n(4000)$ and H$\delta$ parameters. In this case, we are using the MPA-JHU DR7 (\citealt{kau03a}) calculation of the H$\delta$ Lick index, and the D$_n(4000)$ (also from the MPA-JHU catalog), considering the \cite{balogh+99} definition. In both cases, the calculation was done taking into account corrections for emission lines. D$_n(4000)$ is a measurement of the $4000$ {\AA} break, while H$\delta$ is a measurement of the Balmer absorption. The first is related to how long ago the galaxy stopped the star formation (a big break indicates a longer time). This is the case since the break is created due to photon absorption by metals in the stellar atmosphere, and in hot massive stars, the metals are ionized, so the absorption does not happen. The second, H$\delta$, measures bursts of star formation that ended in an interval of $0.1 - 1~Gyr$. After a star formation burst, the galaxy's light is dominated by O and B stars that have weak Balmer absorption. Once their main sequence lifetime is completed, the galaxy light becomes dominated by A-F stars, that have prominent Balmer absorption. It is important to keep in mind that both indexes depend on metallicity, but that is relevant only for old ages (more than $10^9 yr$ after the burst, \citealt{kau03a}). The authors also verified that when those indexes are used together, they provide a powerful probe of the recent star formation history of a galaxy. Hence, we can use both indexes to discriminate between young and old stellar populations. Galaxies with D$_n(4000)$ $< 1.55$ and H$\delta > 2.0$ are characterized by young stellar populations, while the opposite is true for old stellar populations (see \citealt{kau03b}).

In Figure \ref{fig:cdf_dn4000_hdelta} we display the distributions of those two parameters in the bottom panels, while at the top we show their variation with stellar mass. The left panels contain the results regarding D$_n(4000)$, while on the right we have the results for H$\delta$. The analysis is displayed for ETGs, classified as passive, GV, or SF, located in the field and clusters. Independent of the environment, we can see a clear variation from the SF to the Pas population, with the GV in the middle. That suggests the Pas ETGs stopped their star formation longer ago, showing no signs of recent activity. The opposite is true for the SF-ETGs, according to their D$_n(4000)$ and H$\delta$ values. In between, we have the GV population, with residual star formation. On what regards the field {\it vs} cluster comparison, we can see that for the transitional galaxies (GV and SF ETGs) there is a small difference in the D$_n(4000)$ values, but a significant offset in the H$\delta$ {\it vs} M$_*$ plane. That reinforces the scenario of strong(residual) recent activity for the SF(GV) ETGs, which is reduced according to their global environment. However, it is important to notice the field H$\delta$ distributions (bottom right panel, with no mass distinction) always show higher values than the cluster ones. These results are confirmed by the KS-test, (the p-values are listed on the right side of the bottom panel, or in Table \ref{tab:tabela_3}). Regarding the Pas ETG, although there is a significant difference in the distributions in the two environments (field and cluster), they are consistent with quenched galaxies, that stopped their star formation long ago. And, as shown by \cite{kau03b}, for older stellar populations, the metallicity starts to be an important factor. So, for passive galaxies, it is difficult to know if the distinction between the cluster and field samples is driven by age or metallicity (or both).

Another important aspect we detect in the top panels of Figure \ref{fig:cdf_dn4000_hdelta} is the variation with stellar mass. As pointed out by \citet{kau03b}, there is a strong dependence of D$_n(4000)$ and H$\delta$ with M$_*$, with a clear distinction at M$_* \sim 3 \times 10^{10} ~$M$_{\odot}$. We see that is true in our case, for all the populations shown. However, the effect becomes stronger according to the population, being steeper for the SF-ETGs, especially for H$\delta$. On what regards the environment, we can see that for low mass (Log M$_*/$M$_{\odot} < 10.5$) SF-ETGs, there is no environmental distinction. That appears for the higher mass galaxies. For the most massive SF-ETGs, we actually find that cluster objects have no sign of recent activity (according to H$\delta$), but that is not the case for the field counterparts.

We have also investigated the distributions of the color gradient (provided by the KIAS catalog $-$ \citealt{kias}), which gives the color variation inside the galaxy. Although there is a degeneracy between age and metallicity in regard to color, some authors believe that the main driver of the color gradient is the metallicity gradient inside the galaxy (e.g., \citealt{peletier+90, saglia+00, tamura+00a, tamura+00b, kobayashi-04}). In any case, the nature of the formation of the color gradient is not up for debate in the current work, as the important aspect here is the existence of the color gradient \textit{per se}. It is important because we can use the color gradient as a proxy to identify the different types of formation and evolution regimes: monolithic \textit{vs.} hierarchical. In a monolithic scenario, the potential well of the galaxy's central region acts as a retainer of gas. The accumulation of gas in the inner part causes a greater enrichment in the region, leading to a more negative gradient. While, in a hierarchical scenario, due to the mixture and infusion of gas related to merger processes, there is a dilution of the chemical content, causing a less steep gradient \citep{tortora+10, kobayashi-04}. In the case where the age is the main factor in the formation of the color gradient in galaxies, the scenario depicted above is still valid. The center of the galaxy will be older than the outskirts in the monolithic scenario, and in the hierarchical scenario, the infusion of new gas will lead to new star-formation, equalizing the color gradient in the same manner we observe in the case where metallicity is the main factor for the color gradient formation.    

Figure \ref{fig:color_grad_distribution} displays the distribution of the color gradient for our ETG sample. As in other figures, the continuous lines give the distribution of cluster samples, while the dashed lines show the distribution of field samples. The red tones represent the Pas-ETGs, the green tones the GV-ETGs, and the orange tones the SF-ETGs. This figure shows that GV and SF-ETGs have systematically more negative values of $\nabla(g - i)$ than Pas-ETGs, for both environments.

From Figure \ref{fig:color_grad_distribution} it is also possible to compare the different ETGs subpopulations according to the environment, cluster \textit{vs.} field.  In all cases, the distribution of $\nabla(g - i)$ for field galaxies displays systematically more negative values than for cluster galaxies. This result indicates that the environment in which galaxies inhabit acts as a regulator of their evolutionary paths. Field ETGs are more likely to experience monolithic evolution, while cluster ETGs probably have their evolution explained by the hierarchical scenario. Of course, the evolution probably does not follow exclusively one of the scenarios, being more likely that the evolution of galaxies needs both of them to explain all of their properties. However, the distinction in the color gradient can indicate a higher likelihood that the galaxies in a more recent past had been evolving monotonically or hierarchically.

The same trends demonstrated in Figure \ref{fig:color_grad_distribution} for ETGs are observed for LTGs. The comparison of $\nabla$(g - i) between cluster and field LTGs is displayed in Table \ref{tab:tabela_3}. As for the ETGs, the GV and SF-LTGs (and the combined sample) have more negative color gradient values than Pas-LTGs. We also found that field LTGs have more negative values than their cluster counterparts.

In Figure \ref{fig:hist_color_grad} we still show the $\nabla$(g - i) distributions for the ETGs inside clusters and in the field. However, now we distinguish between elliptical (black lines) and lenticular (gray lines) galaxies. For Pas-ETGs in clusters, the distributions of ellipticals and S0s are very similar, displaying only a small difference in the median value of the distributions. A similar behavior is observed for the Pas-ETGs in the Field, but the median values of the distributions show a slightly larger difference. For the GV, SF, and GV+SF-ETGs, the scenario is quite different. For each of these subpopulations, the lenticulars and elliptical galaxies show distinct distributions, having different peaks. In the field, the separation is even larger. 

Figure \ref{fig:hist_color_grad} demonstrates that lenticular galaxies are responsible for the systematical shift in the distribution of $\nabla$(g - i) or the GV, SF, and GV+SF-ETGs subpopulations in relation to the Pas-ETGs. It also shows that most of the shift observed in the comparison between cluster and field ETGs is due to the lenticular galaxies having a stronger tail in the field compared to clusters. That indicates that the likelihood of elliptical and lenticular galaxies following a similar evolutionary path in both environments is small. 

In Figure \ref{fig:cdf_A_and_S} we follow up the discussion of hierarchical \textit{vs.} monolithic scenario observed in Figures \ref{fig:color_grad_distribution} and \ref{fig:hist_color_grad}, by comparing the distributions of Asymmetry (A) and Smoothness (S) of our sub-samples (measured by \citealt{barchi+20}). A is a measurement of how symmetrical a galaxy is, in general, spiral galaxies present higher values of asymmetries than elliptical galaxies \citep{conselice03}. Recent merger remnants may (especially major mergers) present the highest values of asymmetries \citep{palmese+17}, and the value would decrease with time, while the merger remnant reaches a new dynamical equilibrium. On the other hand, smoothness, S is a measurement of how much of the galaxy’s light is contained in small clumps \citep{conselice03}. Therefore, smoothness is higher for unperturbed early-type galaxies. Figure \ref{fig:cdf_A_and_S} shows the CDFs of these two properties, comparing the cluster (solid lines) and the field (dashed lines) samples. The red tones are for Pas-ETGs, the green tones are for GV-ETGs, the orange tones are for SF-ETGs, and the salmon tones are for GV+SF-ETGs. Figure \ref{fig:cdf_A_and_S} indicates a subtle but statistically significant difference in the value of A between cluster and field ETGs, for both passive and star-forming ones - confirmed by the KS and the Anderson-Darling tests in a $95\%$ confidence level - but not for the GV-ETGs. For this case, the null hypothesis is accepted. That is also true for S. These results indicate that cluster ETGs are more asymmetrical and have a less smooth light distribution than field objects.

Due to the nature of the hierarchical formation, it is expected that the galaxies will tend to be more asymmetric and have the light more concentrated in small pockets across the galaxy (consequently, less smooth). This will cause higher values of A and lower values of S. Thus, the results observed in Figure \ref{fig:cdf_A_and_S} partially support the scenario where cluster early-type galaxies have a higher likelihood of having followed a hierarchical formation scenario in comparison to those in the field by showing that cluster ETGs are more asymmetrical and less smooth (i.e., more clumpy). 

We reinforce here that these statements are purely based on galaxy morphological parameters, which might be biased by the object's apparent magnitude (i.e. mass) and apparent size, as well as being affected by the presence of companions (i.e. the environment they live in). More detailed galaxy maps are needed to put stronger constraints on the galaxy's evolutionary paths and its relation with the environment. In a follow-up study, we intend to apply this study to IFU data, to retrieve the spatially resolved star-formation history.

To further attest the cluster environmental influence on galaxy evolution, we display in Figure \ref{fig:sf_frac} the variation of the fractions of star-forming galaxies with normalized clustercentric distance, up to 5$\times$R$_{200}$. As mentioned in \S\ref{sec:data} we have 9931 cluster members within 5$\times$R$_{200}$. In this case, we consider as SF all galaxies above the Eq. \ref{eq:equation2} that separate the passive from the green-valley galaxies in Figure \ref{fig:sfr_vs_mass}. The black continuous curve describes the fractional variation of all star-forming galaxies without morphological distinction. The purple dashed curve displays the results for the SF-LTGs, while the orange dashed curve shows the SF-ETGs sample. The field fraction (for each case) is displayed by the squares on the right side of the figure, for comparison. We can see a positive variation of the fraction of star-forming cluster galaxies up to $\sim 3 \times $R$_{200}$ (reaching a plateau to larger distances). As the SF fractions within clusters are always smaller than the field fractions (even when we go to large distances), these results suggest that galaxies are pre-processed in smaller units (groups) before being accreted by large clusters \citep{haines+15}.

%------------------------------------------------------------------------------------------------------------------%
%------------------------------------------------------------------------------------------------------------------%

\section{Discussion and Conclusions}
\label{sec:discussion_and_conclusions}

One of the main goals of the present work is to use transitional galaxies, \textbf{such as GV and }SF-ETGs as a proxy for galaxies migrating from the BC to the RS and investigate the influence of different environments in this transition. In general, we expect early-type galaxies to be red and dead objects. Nonetheless, several works have been showing that this is not the whole truth. The presence of SF-ETGs objects needs to be accounted for \citep{lee+06, kannappan+09, schawinski+09}. To do so, we have analyzed a sample of $3,899$ cluster galaxies and $11,460$ field galaxies in the local Universe. We further divided the cluster sample between galaxies belonging to relaxed clusters ($2,878$ galaxies) and non-relaxed clusters ($1,021$ galaxies).

Hence, a proper understanding of these populations can shed some light on the evolutionary path of galaxies. According to \cite{vulcani+15}, SF-ETGs comprise about 6$\%$ of the general field ETG population. \cite{kaviraj+11} point out that \textit{ETGs account for around half the stellar mass budget in the local Universe}. They also state that SF-ETGs contribute to $\sim$14 percent of the cosmic star-formation budget.

We have shown in Figure \ref{fig:pps_etg} that the SF-ETGs arrived at the cluster in a later epoch than  Pas-ETGs (for both relaxed and non-relaxed clusters). This is also true for the GV-ETGs and for the combined GV+SF-ETG sample. They also have an infall period similar to the one observed in the late-type population (star-forming or quenched). While the GV and SF-ETGs are dominated by galaxies with recent or intermediate time since infall (same for the LTGs), the Pas-ETGs are dominated by galaxies with ancient infall time. This behavior is consistent with other results in the literature \citep{Jaffe+15, Jaffe+16, lotz+19}, that indicate that the interstellar gas of the galaxies is removed after the first passage into the center of the cluster, preventing further star-formation activity to happens.

While the segregation in the PPS according to the star-formation activity is clear for ETGs, that is not the case for LTGs. Figure \ref{fig:pps_ltg} indicates small differences between the location of Pas, GV, SF, and GV+SF-LTGs in the PPS. We find the fractions of galaxies inside the E region are higher for Pas-LTGs, in comparison to what was observed for GV-LTGs, for the SF-LTGs, and for the GV+SF-ETGs (Figure \ref{fig:pps_etg}). As we saw for the ETGs, this result indicates that the Pas-LTGs may have been within clusters for slightly longer periods than the other LTG subpopulations.

We also show in \S\ref{phase_space+dynamical_state} that the dynamical state of the cluster does not have a major influence on the properties studied in this work (Table \ref{tab:tabela_3}). This result agrees with the findings of \cite{sampaio+21}, that using a different methodology to classify the dynamical state of clusters with M$_{200} > 10^{14}M_\odot$, found that the investigated properties are globally similar in both environments. Although, the galaxies' infall rate of non-relaxed clusters is larger than the one for relaxed clusters.

Unlike what was obtained in \S\ref{phase_space+dynamical_state}, we have shown in the subsection \ref{cluster_vs_field} (comparison of cluster and field galaxies) that the cluster environment influences most of the galaxies' properties. As shown in Table \ref{tab:tabela_3}, for LTGs the scenario is very clear. With exception to the H$\delta$, all the comparisons made in Table \ref{tab:tabela_3} rejected the null hypothesis that the samples were drawn from the same distribution. On the other hand, for the ETGs, the scenarios are more complex, especially for the D$_n(4000)$ and for the sSFR. According to the KS test, the GV and the SF-ETGs from the cluster and the field are drawn from the same distribution, meaning that stellar population age and the star-formation taking place at the cluster and field GV/SF-ETGs are similar. However, when the comparison considers the combined sample (GV+SF-ETGs), the scenario changes. For this enlarged sample, all the properties are distinct between cluster and field (except only for stellar mass). The difference is probably mostly due to the reduced number of galaxies in the individual GV and SF-ETGs samples, yielding  low statistics for the individual comparison. This result indicates that the combined cluster sample of GV+SF-ETGs has an older stellar population - given by the D$_n(4000)$, a more relevant recent star-formation (given by H$\delta$), and a more suppressed ongoing star-formation (given by the sSFR). Since their stellar masses are statistically similar, we can't tribute this difference to a mass effect. This result is in line with several works that have pointed out that different phenomena impact infalling galaxies, affecting their star formation activity \citep{haines+13, zinger+18, wang+18, lotz+19}.

In Figure \ref{fig:cdf_dn4000_hdelta}, we show evidence for a combination of the environment and mass action to quench the star formation activity. We see that for SF-ETGs the environment matters above Log M$_*/$M$_{\odot} \sim 10.5$. The recent star formation activity (indicated by H$\delta$) shows different results, between the field and cluster, only above this mass. In general, the SF-ETGs reduce their activity as they grow in mass, but this process is accelerated for cluster galaxies.

Another result suggested by Table \ref{tab:tabela_3} and demonstrated by Figures \ref{fig:color_grad_distribution} and \ref{fig:hist_color_grad}, is the difference in the color gradient distributions, related to the galaxy environment. As previously argued, the color gradient can give clues to the formation scenario followed by a galaxy. We interpret the results from these two figures (\ref{fig:color_grad_distribution} and \ref{fig:hist_color_grad}) as an indication that cluster galaxies show a higher likelihood of having followed the hierarchical formation scenario, while field galaxies are more likely to follow the monolithic scenario. On top of that, the results from Figure \ref{fig:cdf_A_and_S} (regarding the Asymmetry and Smoothness parameters) suggest that pre-processing of galaxies is an important step before they enter the cluster environment. Before a galaxy becomes a cluster member, it could be part of a galaxy group, and as a group member, the galaxy can suffer similar processes to the ones in the cluster environment (e.g.,  starvation, fly-byes, mergers, harassment, and others; \citealt{haines+15}). Due to the high-velocity dispersion, it's less likely that galaxies go through mergers inside clusters (within R$200$). Thus, it is more probable the results we are observing are the result of mergers that took place before the galaxies become cluster members, where the velocity dispersion is lower than the observed in the interior of the cluster \citep{hickson+97}. In our subsequent project, we intend to investigate the regions where such influences occur by extending our investigation to several R$200$. Preliminary confirmation of the pre-processing effect is displayed here, in Figure \ref{fig:sf_frac}.

Figure \ref{fig:hist_color_grad} offers an attempt to answer the reasoning that the field and cluster samples have different distributions of the color gradient, especially for star-forming ETGs. The bulk of the distinction comes from what we classified as lenticular galaxies. According to this classification, lenticular galaxies are pulling the distribution of field samples to a stronger negative color gradient, indicating a preference for monolithic formation. Our result diverges from other recent works in the literature, like the one of \cite{coccato+20, coccato+22}. In both papers, using IFU data, the authors investigate the different formation scenarios that lenticular galaxies can evolve from. Using the kinematic information obtained by the IFU data, they point out the existence of two types of lenticular galaxies: rotationally-supported and pressure-supported ones. They interpreted each of these types as an indication of the formation processes. Rotationally-supported lenticular would have formed through fast processes that consume their gas (like ram-pressure and starvation). On the contrary, pressure-supported lenticular galaxies would have formed through minor-mergers processes that modified their kinematic characteristics. The authors found an environmental dependency for each of the different types of lenticular galaxies: cluster lenticulars are more rotationally-supported, while field lenticulars are more pressure-supported \citep{coccato+20}. Furthermore, as pointed out in \cite{coccato+22}, \textit{"faded spiral" pathway is the most efficient channel to produce S0s, and it becomes more efficient as the mass of the group or cluster or local density of galaxies increases. The merger pathway is also a viable channel, and its efficiency becomes higher with decreasing local density or environment mass.}. It is important to consider that the morphology classification used in \cite{coccato+20, coccato+22} is not the same for the present work. Also, their stellar mass and luminosity intervals are more constrained to higher luminosity/stellar masses. Furthermore, in \cite{coccato+20, coccato+22}, the authors do not distinguish the lenticular galaxies into passive and star-forming galaxies as we did in the present work. This consideration is relevant when comparing both works since, as was demonstrated in Figures \ref{fig:color_grad_distribution} and \ref{fig:hist_color_grad}, the distinction observed for the evolutionary paths (according to the color gradient distributions) is connected to the star-formation level of the galaxies.

The combination of the results obtained with the color gradient (Figures \ref{fig:color_grad_distribution} and \ref{fig:hist_color_grad}), the Asymmetry and Smoothness (Figure \ref{fig:cdf_A_and_S}), and the star-forming/passive fraction variation with clustercentric distance (Figure \ref{fig:sf_frac}) offer strong argumentation to the need for pre-processing of galaxies before they enter the cluster. The same argument was pointed out by \cite{haines+15}, where the authors investigate a sample of star-forming galaxies coming from 30 massive clusters in a redshift interval of $0.15 < z < 0.30$. The authors found that the fraction of star-forming galaxies increases with cluster-centric radii, but remains below the field value even at 3R$_{200}$. This result can not be reproduced by a scenario where star-formation suppression only occurs in infalling field galaxies, justifying the need for the galaxies to suffer pre-processing before the cluster infall. 

As we have shown, the cluster environment has important consequences for the evolution of galaxies, but is not the only one responsible for the differences observed within the galaxy population. The pre-processing of galaxies must be taken into account when studying galaxy evolution. We intend to give sequence to this work by extending the present analysis into the outer parts of clusters. 

%------------------------------------------------------------------------------------------------------------------%
%------------------------------------------------------------------------------------------------------------------%
%------------------------------------------------------------------------------------------------------------------%
\section*{Acknowledgements}

This work would not be possible without the funding of Coordenação de Aperfeiçoamento de Pessoal de Nível Superior (CAPES), which supported DB with a Ph.D. fellowship. PAAL thanks the support of CNPq, grants 433938/2018-8 e 312460/2021-0. ALBR thanks the support of CNPq, grant 316317/2021-7, and FAPESB INFRA PIE 0013/2016. AC acknowledges the FAPERJ grant E$\_$40/2021 - Apoio ao Jovem Pesquisador Fluminense sem vínculo EM ICTS do Estado do RJ - 2021 - E-26/200.607 and 210.371/2022(270993). We acknowledge the anonymous referee for the very helpful suggestions.

\section*{Data Availability}

The data underlying this article will be shared on reasonable request to the corresponding author.
%------------------------------------------------------------------------------------------------------------------%
%------------------------------------------------------------------------------------------------------------------%
%------------------------------------------------------------------------------------------------------------------%
%%%%%%%%%%%%%%%%%%%%%%%%%%%%%%%%%%%%%%%%%%%%%%%%%%



\bibliographystyle{mnras}
%\bibliography{biblio} % if your bibtex file is called example.bib

% % Alternatively you could enter them by hand, like this:
% % This method is tedious and prone to error if you have lots of references
% \begin{thebibliography}{99}
% \bibitem[\protect\citeauthoryear{Author}{2012}]{Author2012}
% Author A.~N., 2013, Journal of Improbable Astronomy, 1, 1
% \bibitem[\protect\citeauthoryear{Others}{2013}]{Others2013}
% Others S., 2012, Journal of Interesting Stuff, 17, 198
% \end{thebibliography}

%%%%%%%%%%%%%%%%%%%%%%%%%%%%%%%%%%%%%%%%%%%%%%%%%%

%%%%%%%%%%%%%%%%% APPENDICES %%%%%%%%%%%%%%%%%%%%%

%\appendix



%%%%%%%%%%%%%%%%%%%%%%%%%%%%%%%%%%%%%%%%%%%%%%%%%%

% Don't change these lines
\bsp	% typesetting comment
\label{lastpage}
\end{document}